\documentclass[11pt]{llncs} 

\usepackage{a4wide}
\usepackage{amssymb}
\usepackage{amsmath}
\usepackage{color}
\usepackage{xspace}
\usepackage{enumerate}
\usepackage{tikz}
\usetikzlibrary{shapes,backgrounds,plothandlers,plotmarks,calc,arrows,fadings}
\usepackage{ifthen}

\newcommand{\bs}{\backslash}

\def\PF{\noindent{\em Proof:} }
\def\PFSK{\noindent{\em Proof sketch:} }
\def\PFof{\medskip\noindent{\bf Proof of }}
\def\QED{\mbox{}\hspace*{\fill}{$\Box$}\medskip}

\newtheorem{thm}{Theorem}
\newtheorem{lem}[thm]{Lemma}
\newtheorem{propo}[thm]{Proposition}

\newtheorem{corol}[thm]{Corollary}

\def\indset{independent set}

\def\TS{\mbox{TS}}
\def\TJ{\mbox{TJ}}
\def\tsseq{TS-sequence}
\def\tjseq{TJ-sequence}
\def\tarseq{TAR-sequence}

\def\TSreach{TS-Reachability\xspace}
\def\TJreach{TJ-Reachability\xspace}
\def\tsr{\leftrightarrow_{\mbox{\sc ts}}}
\def\tjr{\leftrightarrow_{\mbox{\sc tj}}}
\def\diam{\mbox{diam}}

\def\dist{\mbox{d}}
\def\md{\mbox{md}}
\def\meas{\phi}

\begin{document}

\title{\Large Reconfiguring Independent Sets in Claw-Free Graphs}
\date{\small \today}
\author{Paul Bonsma\inst{1} \and Marcin Kami\'{n}ski\inst{2}  \and Marcin Wrochna\inst{2}}

\institute{University of Twente, Faculty of EEMCS, PO Box 217, 7500 AE Enschede, the Netherlands. Email: {\tt p.s.bonsma@ewi.utwente.nl}
\and
Uniwersytet Warszawski, Institute of Computer Science, Warsaw, Poland. 
Email: {\tt mjk@mimuw.edu.pl} and {\tt mw290715@students.mimuw.edu.pl}} 

\maketitle

\begin{abstract}
We present a polynomial-time algorithm that, given two independent sets in a claw-free graph $G$, decides whether one can be transformed into the other by a sequence of elementary steps. Each elementary step is to remove a vertex $v$ from the current independent set $S$ 
and to add a new vertex $w$ (not in $S$) such that the result is again an \indset.
We also consider the more restricted model where $v$ and $w$ have to be adjacent.
\end{abstract}

\section{Introduction}

\noindent{\bf Reconfiguration problems}. To obtain a reconfiguration version of an algorithmic problem, one defines 
a {\em reconfiguration rule} -- a (symmetric) {adjacency relation} between solutions of the problem, describing small transformations one is allowed to make. The main focus is on studying whether one given solution can be transformed into another by a sequence of such small steps. We call it a {\em reachability problem}.
For example, in a well-studied reconfiguration version of vertex coloring~\cite{BonsmaC09,CerecedaHJ08,CerecedaHJ09,bonamy2011diameter,bonamy2013recoloring,bonamy2014reconfiguration,ito2012reconfiguration}, we are given two $k$-colorings of the vertices of a graph and we should decide whether one can be transformed into the other by recoloring one vertex at a time so that all intermediate solutions are also proper $k$-colorings. 

A useful way to look at reconfiguration problems is through the concept of the \emph{solution graph}. Given a problem instance, the vertices of the {solution graph} are all solutions to the instance, and the reconfiguration rule defines its edges.
Clearly, one
solution can be transformed 
into another if they belong to the same connected 
component of the 
solution graph. 
Other well-studied questions in the context of reconfiguration are as follows: can one efficiently decide (for every instance) whether the solution graph is connected? 
Can one efficiently find shortest paths between two solutions? 
Common non-algorithmic results are giving upper and lower bounds on the possible diameter of components of the solution graph, in terms of the instance size, or studying how much the solution space needs to be increased 
in order to guarantee connectivity.

Reconfiguration is a natural setting for real-life problems in which solutions evolve over time and an interesting theoretical framework that has been gradually attracting more attention. The theoretical interest is  based on the fact that reconfiguration problems  provide a new perspective and offer a deeper understanding of the solution space as well as a potential to develop heuristics to navigate that space. 

Reconfiguration paradigm has been recently applied to a number of algorithmic problems: vertex coloring~\cite{BonsmaC09,BonsmaCHJ07,CerecedaHJ09,CerecedaHJ08}, list-edge coloring \cite{ItoKD09}, clique, set cover, integer programming, matching, spanning tree, matroid bases~\cite{ItoDHPSUU08}, block puzzles \cite{HearnD05}, satisfiability \cite{GopalanKMP09}, independent set \cite{HearnD05,ItoDHPSUU08,KaminskiMM12}, shortest paths \cite{Bonsma12,Bonsma13,KaminskiMM11}, and dominating set \cite{SuzukiMN14}; recently also in the setting of parameterized complexity \cite{MouawadNRSS13}. A recent survey \cite{Heuvel13} gives a good introduction to this area of research. 
\medskip

\noindent{\bf Reconfiguration of independent sets}. The topic of this paper is reconfiguration of independent sets. An {\em independent set} in a graph is a set of pairwise nonadjacent vertices. We will view the elements of an independent set as tokens placed on vertices. Three different reconfiguration rules have been studied in the literature: token sliding (TS), token jumping (TJ), and token addition/removal (TAR). 
The reconfiguration rule in the TS model allows to slide a token along an edge.
The reconfiguration rule in the TJ model allows to remove a token from a vertex and place it on another unoccupied vertex.
In the TAR model, 
the reconfiguration rule allows to either add or remove a token 
as long as at least $k$ tokens remain on the graph at any point, for a given integer $k$. 
In all three cases, the reconfiguration rule may of course only be applied if it maintains an \indset. A sequence of moves following these rules is called a {\em \tsseq, \tjseq, or $k$-\tarseq}, respectively. Note that the TS model is more restricted than the TJ model, in the sense that any \tsseq\ is also a \tjseq. Kami\'nski et al.~\cite{KaminskiMM12} showed that the TAR model generalizes the TJ model, in the sense that there exists a \tjseq\ between two solutions $I$ and $J$ with $|I|=|J|$ if and only if there exists a $k$-\tarseq\ between them, with $k=|I|-1$.
TS seems to have been introduced by Hearn and Demaine~\cite{HearnD05}, 
TAR was introduced by Ito et al.~\cite{ItoDHPSUU11} and TJ by Kami\'nski et al.~\cite{IWOCA2010}.

In all three models, the corresponding reachability problems are {\sf PSPACE}-complete in general graphs~\cite{ItoDHPSUU11} and even in perfect graphs~\cite{KaminskiMM12} or in 
planar graphs of maximum degree~$3$~\cite{HearnD05} 
(see also~\cite{BonsmaC09}). We remark that in~\cite{HearnD05}, only the TS-model was explicitly considered, but since only maximum \indset s are used, this implies the result for the TJ model (see Proposition~\ref{propo:max:TSequivTJ} below) and for the TAR model (using the aforementioned result from~\cite{KaminskiMM12}). 
\medskip

\noindent{\bf Claw-free graphs}. A \emph{claw} is the tree with four vertices and three leaves. A~graph is \emph{claw-free} if it does not contain a claw as an induced subgraph. A claw is not a line graph of any graph and thus the class of claw-free graphs generalizes the class of line graphs. The structure of claw-free graphs is not simple but has  been recently described by Chudnovsky and Seymour in the form of a decomposition theorem \cite{ChudnovskyS05}.

There is a natural one-to-one correspondence between matchings in a graph and independent sets in its line graph. In particular, a maximum matching in a graph corresponds to a maximum independent set in its line graph. Hence, Edmonds' maximum matching algorithm \cite{Edmonds1965a} gives a polynomial-time algorithm for finding maximum independent sets in line graphs. This results has been extended to claw-free graphs independently by Minty~\cite{Minty80} and Sbihi~\cite{sbihi1980algorithme}. Both algorithms work for the unweighted case, while the algorithm of Minty, with a correction proposed by Nakamura and Tamura in \cite{nakamura2001revision}, 
applies to weighted graphs (see also~\cite[Section~69]{schrijver2003combinatorial}). 

A \emph{fork} is the graph obtained from the claw by subdividing one edge. Every claw-free graph is also fork-free. Milani\v c and Lozin gave a polynomial-time algorithm for maximum weighted independent set in fork-free graphs~\cite{LozinM08}. This generalizes all aforementioned results for claw-free graphs. \medskip

\noindent{\bf Our results}. In this paper, we study the reachability problem for independent set reconfiguration, using the TS and TJ model.
Our main result is that these problems can be solved in polynomial time for the case of claw-free graphs.
Along the way, we prove some results that are interesting in their own right. For instance, we show that for connected claw-free graphs, the existence of a \tjseq\ implies the existence of a \tsseq\ between the same pair of solutions. This implies that for connected claw-free and even-hole-free graphs, the solution graph is always connected,
answering an open question posed in~\cite{KaminskiMM12}.

Since claw-free graphs generalize line graphs, our results generalize the result by Ito et al.~\cite{ItoDHPSUU11} on matching reconfiguration. Since a vertex set $I$ of a graph $G$ is an \indset\ if and only if $V(G)\bs I$ is a vertex cover, our results also apply to the recently studied vertex cover reconfiguration problem~\cite{MouawadNRSS13}. The new techniques we introduce can be seen as an extension of the techniques introduced for finding maximum \indset s in claw-free graphs, and we expect them to be useful for addressing similar reconfiguration questions, such as efficiently deciding whether the solution graph is connected. 
\medskip

Some proof details are omitted. Statements for which further details can be found in the appendix are marked with a star.

\section{Preliminaries}
\label{sec:Preliminaries}

For graph theoretic terminology not defined here, we refer to~\cite{Diestel}.
For a graph $G$ and vertex set $S\subseteq V(G)$, we denote the subgraph induced by $S$ by $G[S]$, and denote $G-S=G[V\bs S]$. The set of neighbors of a vertex $v\in V(G)$ is denoted by $N(v)$, and the closed neighborhood of $v$ is $N[v]=N(v)\cup \{v\}$. A {\em walk} from $v_0$ to $v_k$ of length $k$ is a sequence of vertices $v_0,v_1,\ldots,v_k$ such that $v_iv_{i+1}\in E(G)$ for all $i\in \{0,\ldots,k-1\}$. It is a {\em path} if all of its vertices are distinct, and a {\em cycle} if $k\ge 3$, $v_0=v_k$ and $v_0,\ldots,v_{k-1}$ is a path.
We use $V(C)$ to denote the vertex set of a path or cycle, viewed as a subgraph of $G$. A path or graph is called {\em trivial} if it contains only one vertex.
Edges of a directed graph or {\em digraph} $D$ are called {\em arcs}, and are denoted by the ordered tuple $(u,v)$. A {\em directed path} in $D$ is a sequence of distinct vertices $v_0,\ldots,v_k$ such that for all $i\in\{0,\ldots,k-1\}$, $(v_i,v_{i+1})$ is an arc of $D$.

We denote the distance of two vertices $u,v\in V(G)$ by $\dist_G(u,v)$. 
By $\diam(G)$ we denote the {\em diameter} of a connected graph $G$, defined as $\max_{u,v\in V(G)} \dist_G(u,v)$. 
For a vertex set $S$ of a graph $G$ and integer $i\in \mathbb{N}$, we denote
$N_i(S)=\{v\in V(G)\bs S : |N(v)\cap S|=i\}$.

For a graph $G$, by $\TS_k(G)$ we denote the graph that has as its vertex the set of all independent sets of $G$ of size $k$, where two independent sets $I$ and $J$ are adjacent if there is an edge $uv\in E(G)$ with $I\bs J=\{u\}$ and $J\bs I=\{v\}$. We say that $J$ can be obtained from $I$ by {\em sliding a token from $u$ to $v$}, or by the {\em move $u\to v$} for short. A walk in $\TS_k(G)$ from $I$ to $J$ is called a {\em \tsseq\ from $I$ to $J$}. We write $I\tsr J$ to indicate that there is a \tsseq\ from $I$ to $J$.

Analogously, by $\TJ_k(G)$ we denote the graph that has as its vertex set the set of all independent sets of $G$ of size $k$, where two independent sets $I$ and $J$ are adjacent if there is a vertex pair $u,v\in V(G)$ with $I\bs J=\{u\}$ and $J\bs I=\{v\}$. We say that $J$ can be obtained from $I$ by {\em jumping a token from $u$ to $v$}. A walk in $\TS_k(G)$ from $I$ to $J$ is called a {\em \tjseq\ from $I$ to $J$}. We write $I\tjr J$ to indicate that there exists a \tjseq\ from $I$ to $J$.
Note that $\TS_k(G)$ is a spanning subgraph of $\TJ_k(G)$.
 
The {\em reachability problem} for token sliding (resp.\ token jumping) has as input a graph $G$ and two independent sets $I$ and $J$ of $G$ with $|I|=|J|$, and asks whether $I\tsr J$ (resp.\ $I\tjr J$). These problems are called \TSreach and \TJreach, respectively.

If $H$ is a claw with vertex set $\{u,v,w,x\}$ such that $N(u)=\{v,w,x\}$, then $H$ is called a {\em $u$-claw with leaves $v,w,x$.} Sets $I\bs\{v\}$ and $I\cup\{v\}$ are denoted by $I-v$ and $I+v$ respectively. The symmetric difference of two sets $I$ and $J$ is denoted by $I\Delta J=(I\bs J)\cup (J\bs I)$. The following observation is used implicitly in many proofs:
if $I$ and $J$ are independent sets in a claw-free graph $G$, then every component of $G[I\Delta J]$ is a path or an even length cycle.

By $\alpha(G)$ we denote the size of the largest independent set of $G$. An independent set $I$ is called {\em maximum} if $|I|=\alpha(G)$. 
A vertex set $S\subseteq V(G)$ is a {\em dominating set} if $N[v]\cap S\not=\emptyset$ for all $v\in V(G)$. Observe that a maximum independent set is a dominating set, thus the only possible token jumps from it are between adjacent vertices, and hence all are token slides:

\begin{propo}
\label{propo:max:TSequivTJ} 
Let $I$ and $J$ be maximum independent sets in a  graph $G$. Then, $TS_k(G) = TJ_k(G)$. In particular, $I \tsr J$ if and only if $I\tjr J$.
\end{propo}

\section{The Equivalence of Sliding and Jumping}
\label{sec:TSvsTJ}

In our main result (Theorem~\ref{thm:connected_reachability}), we will consider equal size \indset s $I$ and $J$ of a claw-free graph $G$, and show that in polynomial time, it can be verified whether $I\tsr J$ and whether $I\tjr J$. In this section, we show that if $G$ is connected and $G[I\Delta J]$ contains no cycles, then $I\tsr J$. From this, we will subsequently conclude that for connected claw-free graphs  $I\tsr J$ holds if and only if $I\tjr J$, even in the case of nonmaximum independent sets.

\begin{lem}[*]
\label{lem:NoBadCyclesSliding}
Let $I$ and $J$ be independent sets in a connected claw-free graph $G$ with $|I|=|J|$. If $G[I\Delta J]$ contains no cycles, then $I\tsr J$.
\end{lem}

\PFSK
We show that $I$ or $J$ can be modified using token slides such that the two resulting \indset s are closer to each other in the sense that either $|I\setminus J|$ is smaller, or it is unchanged and the minimum distance between vertices $u,v$ with $u\in I\setminus J$ and $v\in J\setminus I$ is smaller. The claim follows by induction. (See the appendix for an induction statement with a bound on the length of the reconfiguration sequence.)

Suppose first that $G[I\Delta J]$ contains at least one nontrivial component $C$. Since it is not a cycle by assumption, it must be a path. Choose an end vertex $u$ of this path, and let $v$ be its unique neighbor on the path. If $u\in J$ then $N(u)\cap I=\{v\}$, so we can obtain a new \indset\ $I'=I+u-v$ from $I$ using a single token slide. The new set $I'$ is closer to $J$ in the sense that $|I'\bs J|<|I\bs J|$, 
so we may use induction to conclude that
$I'\tsr J$, and thus $I\tsr J$. 
On the other hand, if $u\in I$ then we can obtain a new \indset\ $J'=J-v+u$ from $J$, and 
conclude the proof similarly by applying the induction assumption to $J'$ and $I$.

In the remaining case, we may assume that $G[I\Delta J]$ consists only of isolated vertices. 
Choose $u\in I\bs J$ and $v\in J\bs I$, such that the distance $d:=\dist_G(u,v)$ between these vertices is minimized. Starting with $I$, we intend to slide the token on $u$ to $v$, to obtain an \indset\ $I'=I-u+v$ that is closer to $J$. 
To this end, we choose a shortest path $P=v_0,\ldots,v_d$ in $G$ from $v_0=u$ to $v_d=v$. If the token can be moved along this path while maintaining an \indset\ throughout, then $I\tsr I'$, and the proof 
follows by induction as before.

So now suppose that this cannot be done, that is, at least one of the vertices on $P$ is equal to or adjacent to a vertex in $I-u$. In that case, we choose $i$ maximum such that $N(v_i)\cap I\not=\emptyset$.
Using some simple observations (including the fact that $G$ is claw-free), one can now show that $N(v_i)\cap I$ consists of a single vertex $x$. By choice of $v_i$, starting with $I$, the token on $x$ can be moved along the path $x,v_i,v_{i+1},\ldots,v_d$ while maintaining an \indset\ throughout. This yields an \indset\ $I''=I-x+v$, with $I\tsr I''$. It can also easily be shown that $\dist_G(u,x)<\dist_G(u,v)$ and $\dist_G(x,v)<\dist_G(u,v)$. So considering the choice of $u$ and $v$, it follows that $x\in I\cap J$, and thus $|I''\bs J|=|I\bs J|$. Since now the pair $u\in I''\bs J$ and $x\in J\bs I''$ has a smaller distance $\dist_G(u,x)<\dist_G(u,v)=d$, we may assume by induction that $I''\tsr J$, and thus $I\tsr J$.\QED

\begin{corol}
\label{cor:connected:TSequivTJ}
Let $I$ and $J$ be independent sets in a connected claw-free graph $G$. Then $I \tsr J$ if and only if $I\tjr J$.
\end{corol}

\PF
Clearly, a \tsseq\ from $I$ to $J$ is also a \tjseq. For the nontrivial direction of the proof, it suffices to show that any token jump can be replaced by a sequence of token slides. Let $J$ be obtained from $I$ by jumping a token from $u$ to $v$. Then $G[I\Delta J]$ contains only two vertices and therefore no cycles. Then Lemma~\ref{lem:NoBadCyclesSliding} shows that $I\tsr J$.
\QED

We now consider implications of the above corollary for graphs that are claw- and even-hole-free. A graph is {\em even-hole-free} if it contains no even cycle as an induced subgraph. Kami\'{n}ski et al.~\cite{KaminskiMM12} proved the following statement.
\begin{thm}[\cite{KaminskiMM12}]
\label{thm:KMM}
Let $I$ and $J$ be two \indset s of a graph $G$ with $|I|=|J|$. If $G[I\Delta J]$ contains no even cycles, then there exists a \tjseq\ from $I$ to $J$ of length $|I\bs J|$, which can be constructed in linear time. 
\end{thm}
In particular, if $G$ is even-hole-free, then $\TJ_k(G)$ is connected (for every $k$). However, $\TS_k(G)$ is not necessarily connected (consider a claw with two tokens). This motivated the question asked in~\cite{KaminskiMM12} whether for connected, claw-free and even-hole-free graph $G$, $\TS_k(G)$ is connected. 
Combining Corollary~\ref{cor:connected:TSequivTJ} with Theorem~\ref{thm:KMM} shows that the answer to this question is affirmative.

\begin{corol}
Let $G$ be a connected claw-free and even-hole-free graph. Then $\TS_k(G)$ is connected. 
\end{corol}

\section{Nonmaximum Independent Sets}
\label{sec:nonmaximum}

We now continue studying connected claw-free graphs. Lemma~\ref{lem:NoBadCyclesSliding} shows that it remains to consider the case that  $G[I\Delta J]$ contains (even length) cycles. In this section, we show that when $I$ and $J$ are not maximum \indset s of $G$, such cycles can always be resolved. This requires various techniques developed in the context of finding maximum \indset s in claw-free graphs and the following definitions.

A vertex $v\in V(G)$ is {\em free} (with respect to an independent set $I$ of $G$) if 
$v \notin I$ and $|N(v)\cap I|\le 1$.
Let $W=v_0,\ldots,v_k$ be a walk in $G$, and let $I\subseteq V(G)$. Then $W$ is called {\em $I$-alternating} if $|\{v_i,v_{i+1}\}\cap I|=1$ for $i=0,\dots,k-1$.
In the case that $W$ is a path, $W$ is called {\em chordless} if $G[\{v_0,\ldots,v_k\}]$ is a path. In the case that $W$ is a cycle (so $v_0=v_k$), $W$ is called {\em chordless} if $G[\{v_0,\ldots,v_{k-1}\}]$ is a cycle.
A cycle $W=v_0,\ldots,v_k$ is called {\em $I$-bad} if
it is $I$-alternating and chordless. 
A path $W=v_0,\ldots,v_k$ with $k\ge 2$ is called {\em $I$-augmenting} if
it is $I$-alternating and chordless, and
$v_0$ and $v_k$ are both free vertices.
This definition of $I$-augmenting paths differs from the usual definition, as it is used in the setting of finding {\em maximum independent sets}, since the chordless condition is stronger than needed in such a setting. 
However, we observe that in a claw-free graph $G$, the two definitions are equivalent (see Proposition~\ref{propo:AugmPathDefEquiv} in the appendix) so we may apply well-known statements about $I$-augmenting paths proved elsewhere. In particular, we use the following two results originally proved by Minty~\cite{Minty80} and Sbihi~\cite{sbihi1980algorithme} (see also~\cite[Section~69.2]{schrijver2003combinatorial}).

\begin{thm}[\cite{schrijver2003combinatorial}]
\label{thm:AugmPathPoly}
Let $I$ be an \indset\ in a claw-free graph $G$. It can be decided in polynomial time whether an $I$-augmenting path between two given free vertices $x$ and $y$ exists, and if so, compute one.
\end{thm}

\begin{propo}[\cite{schrijver2003combinatorial}]
\label{propo:NotMaxImpliesAugmentable}
Let $I$ be a nonmaximum independent set in a claw-free graph $G$. 
Then $I$ is not a dominating set, or there exists an $I$-augmenting path. 
\end{propo}

We use Proposition~\ref{propo:NotMaxImpliesAugmentable} to handle the case of nonmaximum \indset s. The next statement is formulated for token jumping, and (by Corollary~\ref{cor:connected:TSequivTJ}) implies the same result for token sliding only in the case of connected graphs.

\begin{lem}[*]
\label{lem:non_maximum_NEW}
Let $I$ be a nonmaximum independent set in a claw-free graph $G$. Then for any independent set $J$ with $|J|=|I|$, $I\tjr J$ holds.
\end{lem}

\PFSK
By 
Theorem~\ref{thm:KMM}, 
it suffices to consider the case where $G[I\Delta J]$ contains at least one cycle $C$. 
Let $C=u_1,v_1,u_2,v_2,\ldots,v_k,u_1$, so that $u_i\in I$ and $v_i\in J$ for all $i$.

Suppose first that $I$ is not a dominating set. Then we can choose a vertex $w$ with $N[w]\cap I=\emptyset$.
With a single token jump, we can obtain the \indset\ $I'=I+w-u_1$ from $I$. Next, apply the moves $u_k\to v_k$, $u_{k-1}\to v_{k-1}$,\ldots, $u_2\to v_2$, in this order. (This is possible since $C$ is chordless.) Finally, jump the token from $w$ to $v_1$. It can be verified that this yields a token jumping sequence from $I$ to $I'=I\Delta V(C)$. This way, all cycles can be resolved one by one, until no more cycles remain and 
Theorem~\ref{thm:KMM} 
can be applied to prove the statement.

On the other hand, if $I$ is a dominating set, then Proposition~\ref{propo:NotMaxImpliesAugmentable} shows that there exists an $I$-augmenting path $P=v_0,u_1,v_1,\ldots,u_d,v_d$, with $u_i\in I$ for all $i$. Since $v_d$ is a free vertex, we can first apply the moves $u_d\to v_d$, $u_{d-1}\to v_{d-1}$,\ldots $u_1\to v_1$, in this order (which can be done since $P$ is chordless), to obtain an \indset\ $I'$ from $I$, with $I\tsr I'$. 
Then $v_0$ is not dominated by $I'$, so the previous argument can be applied to show that $I'\tjr J$, which implies $I\tjr J$.
\QED

\section{Resolving Cycles}
\label{sec:resolving}

It now remains to study the case where $G[I\Delta J]$ contains (even) cycles and both $I$ and $J$ are maximum independent sets. 
In this case, there may not be a \tsseq\ from $I$ to $J$ (even though we assume that $G$ is connected and claw-free) -- consider for instance the case where $G$ itself is an even cycle. In this section, we characterize the case where $I\tsr J$ holds, by showing that this is equivalent with every cycle being resolvable in a certain sense (Theorem~\ref{thm:reachable_iff_all_resolvable} below).
Subsequently, we show that resolvable cycles fall into two cases: internally or externally resolvable cycles, which are characterized next. We first define the notion of resolving a cycle. 

Cycles in $G[I\Delta J]$ are clearly 
both $I$-bad and $J$-bad. 
The {\em $I$-bipartition} of an $I$-bad cycle is the ordered tuple $[V(C)\cap I,V(C)\bs I]$.
We say that an $I$-bad cycle $C$ with $I$-bipartition $[A,B]$ is {\em resolvable} (with respect to $I$) if there exists an \indset\ $I'$ such that $I\tsr I'$ and $G[I'\cup B]$ contains no cycles. 
A corresponding \tsseq\ from $I$ to $I'$ is called a {\em resolving sequence} and is said to {\em resolve $C$}. 
By combining such a resolving sequence with a sequence of moves similar to the previous proof, and then reversing the moves in the sequence from $I'$ to $I$, except for moves of tokens on the cycle, one can show that every resolvable cycle can be `turned':

\begin{lem}[*]
\label{lem:resolvable_implies_turnable}
Let $I$ be an independent set in a claw-free graph $G$ and let $C$ be an $I$-bad cycle. If $C$ is resolvable with respect to $I$, then $I\tsr I\Delta V(C)$.
\end{lem}

We can now prove the following useful characterization: $I\tsr J$ if and only if every cycle in $G[I\Delta J]$ is resolvable. By symmetry, it does not matter whether one considers resolvability with respect to $I$ or to $J$. 

\begin{thm}
\label{thm:reachable_iff_all_resolvable}
Let $I$ and $J$ be 
independent sets
in a claw-free connected graph $G$. 
Then $I\tsr J$ if and only if every cycle in $G[I\Delta J]$
is resolvable with respect to $I$.
\end{thm}
\PF 
Consider an $I$-bad cycle $C$ in $G[I\Delta J]$ with $I$-bipartition $[A,B]$, and a \tsseq\ from $I$ to $J$. Since $N_2(B)$ eventually contains no tokens, this sequence must contain a move $u\to v$ with $u\in N_2(B)$ and $v\not\in N_2(B)$. The first such move can be shown to resolve the cycle. (See Lemma~\ref{lem:ShortResolvingSeqNEW} in the appendix for details.)

The other direction is proved by induction on the number $k$ of cycles in $G[I\Delta J]$.
If $k=0$, then by Lemma~\ref{lem:NoBadCyclesSliding}, $I\tsr J$. 
If $k\ge 1$, then consider an $I$-bad cycle $C$ in $G[I\Delta J]$. Let $I'=I\Delta V(C)$. 
By Lemma~\ref{lem:resolvable_implies_turnable}, $I\tsr I'$. The graph $G[I'\Delta J]$ has one cycle fewer than $G[I\Delta J]$. Every cycle in $G[I'\Delta J]$ remains resolvable with respect to $I'$ (one can first consider a \tsseq\ from $I'$ to $I$, and subsequently a \tsseq\ from $I$ that resolves the cycle). So by induction, $I'\tsr J$, and therefore, $I\tsr J$.\QED

Finally, we show that if an $I$-bad cycle $C$ can be resolved, it can be resolved in at least one of two very specific ways. 
Let $[A,B]$ be the $I$-bipartition of $C$.
A move 
$u\to v$ is called {\em internal} if $\{u,v\}\subseteq N_2(B)$ and {\em external} if $\{u,v\}\subseteq N_0(B)$.
A resolving sequence $I_0,\ldots,I_m$ for $C$ is called {\em internal} (or {\em external}) if every move except the last is an internal (respectively, external) move. 
(Obviously, to resolve the cycle, the last move can neither be internal nor external, and can in fact be shown to always be a move from $N_2(B)$ to $N_1(B)$.)
The $I$-bad cycle $C$ is called {\em internally resolvable} resp.\ {\em externally resolvable} if such sequences exist.

\begin{lem}[*]
\label{lem:externally_or_internally}
Let $I$ be an independent set
in a claw-free graph $G$ and let $C$ be an $I$-bad cycle. 
Then 
any shortest \tsseq\ that resolves $C$ is an internal or external resolving sequence.
\end{lem}
\PFSK
Let $[A,B]$ be the $I$-bipartition of $C$.
Since $G$ is claw-free, it follows that there are no edges between vertices in $N_2(B)$ and $N_0(B)$. This can be used to show that informally, any resolving sequence for $C$ remains a resolving sequence after either omitting all noninternal moves or omitting all nonexternal moves, while keeping the last move, which subsequently resolves the cycle.
\QED

Theorem~\ref{thm:reachable_iff_all_resolvable} and Lemma~\ref{lem:externally_or_internally} show that to decide whether $I\tsr J$, it suffices to check whether every cycle in $G[I\Delta J]$ is externally or internally resolvable. 
Next we give characterizations that allow polynomial-time algorithms for deciding whether an $I$-bad cycle is internally or externally resolvable. 
For the external case, we use the assumption that $I$ is a maximum \indset\ to show that in a {\em shortest} external resolving sequence $I_0,\ldots,I_m$, every token moves at most once (that is, for every move $u\to v$, both $u\in I_0$ and $v\in I_m$ hold), so 
these moves outline an augmenting path in a certain auxiliary graph.

\begin{thm}
\label{thm:ext_res_augm_path}[*]
Let $I$ be a maximum \indset\ in a claw-free graph $G$ and let $C$ be an $I$-bad cycle with $I$-bipartition $[A,B]$. Then $C$ is externally resolvable if and only if there exists an $(I\bs A)$-augmenting path in $G-A-B$ between a pair of vertices $x\in N_0(B)$ and $y\in N_1(B)$.
\end{thm}

For a given $I$-bad cycle $C$ with $I$-bipartition $[A,B]$, there is a quadratic number of vertex pairs $x\in N_0(B)$ and $y\in N_1(B)$ that need to be considered, and for every such a pair, testing whether there is an $(I\bs A)$-augmenting path between these in $G-A-B$ can be done in polynomial time (Theorem~\ref{thm:AugmPathPoly}). So from Theorem~\ref{thm:ext_res_augm_path} we conclude:

\begin{corol}
\label{cor:ext_res_polytime} 
Let $I$ be a maximum \indset\ in a claw-free graph $G$, and let $C$ be an $I$-bad cycle. In polynomial time, it can be decided whether $C$ is externally resolvable.
\end{corol}

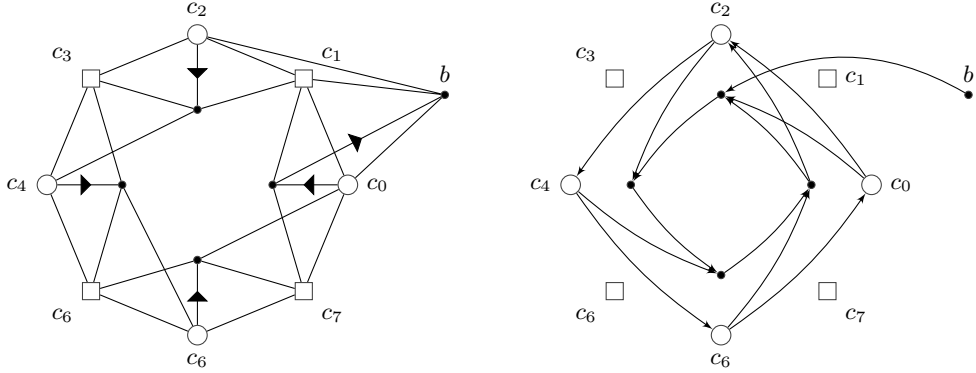
\begin{figure}[t]
\centering
\newcommand{\midarrow}{\tikz \draw[-triangle 90] (0,0) -- +(.1,0);}
\begin{tabular}{c c c}
\begin{tikzpicture}[every node/.style={sloped,allow upside down}]
	\tikzstyle{v}=[circle,fill=black,draw=black!75,inner sep=0pt,minimum size=0.3em]
	\tikzstyle{I}=[circle,draw=black!75,inner sep=0pt,minimum size=0.8em]
	\tikzstyle{J}=[rectangle,draw=black!75,inner sep=0pt,minimum size=0.7em]
	
	\node[I,label=   0:$c_0$] (c0) at (   0:2) {};
	\node[J,label=  45:$c_1$] (c1) at (  45:2) {};	
	\node[I,label=  90:$c_2$] (c2) at (  90:2) {};		
	\node[J,label= 135:$c_3$] (c3) at ( 135:2) {};	
	\node[I,label= 180:$c_4$] (c4) at ( 180:2) {};	
	\node[J,label=-135:$c_6$] (c5) at (-135:2) {};	
	\node[I,label= -90:$c_6$] (c6) at ( -90:2) {};	
	\node[J,label= -45:$c_7$] (c7) at ( -45:2) {};
	\draw (c0)--(c1)--(c2)--(c3)--(c4)--(c5)--(c6)--(c7)--(c0);
	\node[v] (u4) at (0:1) {};
	\node[v,label=$b$] (u0) at (20:3.5) {};	
	\node[v] (u1) at (90:1) {};
	\node[v] (u2) at (180:1) {};
	\node[v] (u3) at (-90:1) {};
	\draw (u4)--(c7) (u4)--(c1); \draw (c0)--node{\midarrow}(u4);
	\draw (u3)--(c0) (u3)--(c7) (c6)--node{\midarrow}(u3) (u3)--(c5);
	\draw (u2)--(c6) (u2)--(c5) (c4)--node{\midarrow}(u2) (u2)--(c3);	
	\draw (u1)--(c4) (u1)--(c3) (c2)--node{\midarrow}(u1) (u1)--(c1);	
	\draw (u0)--(c2) (u0)--(c1) (u0)--(c0) (u4)--node{\midarrow}(u0);
\end{tikzpicture}
& \hspace{2em} &
\begin{tikzpicture}[->,arrows=-latex']
	\tikzstyle{v}=[circle,fill=black,draw=black!75,inner sep=0pt,minimum size=0.3em]
	\tikzstyle{I}=[circle,draw=black!75,inner sep=0pt,minimum size=0.8em]
	\tikzstyle{J}=[rectangle,draw=black!75,inner sep=0pt,minimum size=0.7em]
	\tikzstyle{every path}=[draw,bend right=10] ;
	
	\node[I,label=   0:$c_0$] (c0) at (   0:2) {};
	\node[J,label=   0:$c_1$] (c1) at (  45:2) {};	
	\node[I,label=  90:$c_2$] (c2) at (  90:2) {};		
	\node[J,label= 135:$c_3$] (c3) at ( 135:2) {};	
	\node[I,label= 180:$c_4$] (c4) at ( 180:2) {};	
	\node[J,label=-135:$c_6$] (c5) at (-135:2) {};	
	\node[I,label= -90:$c_6$] (c6) at ( -90:2) {};	
	\node[J,label= -45:$c_7$] (c7) at ( -45:2) {};
	\node[v] (u4) at (0:1.2) {};
	\node[v,label=$b$] (u0) at (20:3.5) {};	
	\node[v] (u1) at (90:1.2) {};
	\node[v] (u2) at (180:1.2) {};
	\node[v] (u3) at (-90:1.2) {};
	\draw[bend right=30] (u0)to(u1);
	\draw (u1)to(u2); \draw (u2)to(u3);
	\draw (u3)to(u4); \draw (u4)to(u1); \draw (u4)to(c2);
	\draw (c2)to(u2); \draw (c2)to(c4);
	\draw (c4)to(u3); \draw (c4)to(c6);	
	\draw (c6)to(u4); \draw (c6)to(c0);
	\draw (c0)to(u1); \draw (c0)to(c2);
\end{tikzpicture}
\end{tabular}

\caption{An example of a claw-free graph $G$ with an internally resolvable cycle, along with the corresponding auxiliary digraph $D(G,C)$.}
\label{fig:internal_example}
\end{figure}

Next, we characterize internally resolvable cycles. Shortest internal resolving sequences cannot be as easy to describe as external ones, since a token can move several times (see Figure~\ref{fig:internal_example}). Nevertheless, these sequences can be shown to have a very specific structure, which can be characterized using paths in the following auxiliary digraphs.

To define these digraphs, 
consider an $I$-bad cycle $C=c_0,c_1,\ldots,c_{2n-1},c_0$ in $G$, with $c_i\in I$ for even $i$. 
Let $[A,B]$ be the $I$-bipartition of $C$.
For every $i\in \{0,\ldots,n-1\}$, define the corresponding {\em layer} as follows: $L_i=\{v\in V(G) \mid N(v)\cap B=N(c_{2i})\cap B\}$. So when starting with $I$ and using only internal moves, it can be seen that the token that starts on $c_{2i}$ will stay in the layer $L_i$.

For such an $I$-bad cycle $C$ of length at least 8, define $D(G,C)$ to be a digraph with vertex set $V(G)$, with the following arc set. 
For every $i\in\{0,\ldots,n-1\}$ and all pairs $u\in L_i, v\in L_{(i+1)\bmod n}$ with $uv\not\in E(G)$, add an arc $(u,v)$. 
For every $i\in\{0,\ldots,n-1\}$ and $b\in N_1(B)$ with $N(b)\cap B=\{c_{(2i-1) \bmod 2n}\}$, and every $v\in L_i$ with $bv\not\in E(G)$, add an arc $(b,v)$. 
Also, we denote the reversed cycle by $C^{rev}=c_0,c_{2n-1},\ldots,c_1,c_0$. This defines a similar digraph $D(G,C^{rev})$ (where arcs between layers are reversed, and arcs from $N_1(B)$ go to different layers). These graphs can be used to characterize whether $C$ is internally resolvable.

\begin{thm}[*]
\label{thm:internally_iff_copath}
Let $I$ be an 
independent set in a claw-free graph $G$. Let $C=c_0,c_1,\dots,c_{2n-1},c_{0}$ be an $I$-bad cycle ($c_0\in I$) with $I$-bipartition $[A,B]$, of length at least 8. 
Then $C$ is internally resolvable if and only if $D(G,C)$ or $D(G,C^{rev})$ contains a directed path from a vertex $b\in N_1(B)$ with $N(b)\cap I\subseteq A$ to a vertex in $A$.
\end{thm}

\begin{corol}
\label{cor:int_res_polytime}
Let $I$ be an independent set in a claw-free graph $G$ on $n$ vertices and let $C$ be an $I$-bad cycle. It can be decided in polynomial time whether $C$ is internally resolvable.
\end{corol}

\PF
If $C$ has length at least 8, then Theorem~\ref{thm:internally_iff_copath} shows that it suffices to make a polynomial number of depth-first-searches in $D(G,C)$ and $D(G,C^{rev})$. Otherwise, let $[A,B]$ be the $I$-bipartition of $C$. $|A|\le 3$, so there are only $O(n^3)$ independent sets $I'$ with 
$|I'|=|I|$ and $I\bs A\subseteq I'$.
So in polynomial time we can generate the subgraph of $\TS_k(G)$ induced by these sets, and search whether it contains a path from $I$ to an \indset\ $I^*$ with $I\bs A\subseteq I^*$ where $G[B\cup I^*]$ contains no cycle. $C$ is internally resolvable if and only if such a path exists. 
\QED

\section{Summary of the Algorithm}
\label{sec:summary}

We now summarize how the previous lemmas yield a polynomial time algorithm for \TSreach and \TJreach in claw-free graphs.

\begin{thm}
\label{thm:connected_reachability}
Let $I$ and $J$ be independent sets in a claw-free graph $G$. We can 
decide in polynomial time whether $I\tsr J$ and whether $I\tjr J$. 
\end{thm}

\PF
Assume $|I|=|J|$; otherwise, we immediately return NO. We first consider the case when $G$ is connected. By Corollary~\ref{cor:connected:TSequivTJ}, since $G$ is connected, $I \tsr J$ if and only if $I\tjr J$, thus we only need to consider the sliding model. 

We test whether $I$ and $J$ are maximum \indset s of $G$, which can be done in polynomial time (by combining Proposition~\ref{propo:NotMaxImpliesAugmentable} and Theorem~\ref{thm:AugmPathPoly}; see also~\cite{Minty80,sbihi1980algorithme,schrijver2003combinatorial}). 
If not, then by Lemma~\ref{lem:non_maximum_NEW}, $I\tjr J$ holds, and thus $I\tsr J$, so we may we return YES.

Now consider the case that both $I$ and $J$ are maximum \indset s. 
Theorem~\ref{thm:reachable_iff_all_resolvable} shows that $I\tsr J$ if and only if every cycle in $G[I\Delta J]$ is resolvable with respect to $I$.
By Lemma~\ref{lem:externally_or_internally}, it suffices to check for internal and external resolvability of such cycles. This can be done in polynomial time by Corollary~\ref{cor:ext_res_polytime} (since $I$ is a maximum \indset\ of $G$) and Corollary~\ref{cor:int_res_polytime}. 
We return YES if and only if every cycle in $C$ was found to be internally or externally resolvable, and NO otherwise.

Now let us consider the case when $G$ is disconnected. Clearly tokens cannot slide between different connected components, so for deciding whether $I\tsr J$, we can apply the argument above to every component, and return YES if and only if the answer is YES for every component. If $I$ is a not a maximum \indset\ then Lemma~\ref{lem:non_maximum_NEW} shows that $I\tjr J$ always holds. If $I$ is maximum, then Proposition~\ref{propo:max:TSequivTJ} shows that $I\tjr J$ holds if and only if $I\tsr J$.
\QED

\section{Discussion}
\label{sec:discussion}

The results presented here have two further implications. Firstly, combined with techniques from~\cite{Bonsma14}, it follows that $I\tjr J$ can be decided for any graph $G$ that can be obtained from a collection of claw-free graphs using {\em disjoint union} and {\em complete join} operations. See~\cite{Bonsma14} for more details. 

Secondly, a closer look at constructed reconfiguration sequences (in the appendix) shows that when $G$ is claw-free, components of both $\TS_k(G)$ and $\TJ_k(G)$ have diameter bounded polynomially in $|V(G)|$. 
This is not surprising, since the same behavior has been observed many times. To our knowledge, the only known examples of polynomial time solvable reconfiguration problems that nevertheless require exponentially long reconfiguration sequences are on artificial instance classes, which are constructed particularly for this purpose (see e.g.~\cite{BonsmaC09,KMM11}).

\appendix

\section{Details for Section~\ref{sec:TSvsTJ}}

We now give a detailed induction proof of Lemma~\ref{lem:NoBadCyclesSliding}, including a bound on the length of the resulting \tsseq.

\begin{lem}
\label{lem:NoBadCyclesSliding_InclDiam}
Let $I$ and $J$ be independent sets in a connected claw-free graph $G$, with $|I|=|J|$. If $G[I\Delta J]$ contains no cycles then there is a \tsseq\ from $I$ to $J$ of length at most $2\cdot |I\bs J|\cdot \diam(G)$.
\end{lem}

\PF
For two vertex sets $I$ and $J$ with $I\bs J\not=\emptyset$ and $J\bs I\not=\emptyset$, define the {\em minimum distance} $\md(I,J)$ to be the minimum of $\dist_G(u,v)$ over all pairs $u\in I\bs J$ and $v\in J\bs I$. In addition, define $\meas(I,J)=(|I\bs J|-1)\cdot \diam(G) + \md(I,J)$.
We will prove by induction on $\meas(I,J)$ that
if $I$ and $J$ are two \indset s with $|I|=|J|$ such that $G[I\Delta J]$ contains no cycles, then there exists a \tsseq\ from $I$ to $J$ of length at most $2\meas(I,J)$.
Since $2\meas(I,J)\leq 2\cdot |I\bs J|\cdot \diam(G)$, this proves the lemma.

First let us consider the case that $\md(I,J)=1$. This means that $G[I\Delta J]$ contains at least one edge. Since $G$ is claw-free, $G[I\Delta J]$ has maximum degree 2. But we assumed that it contains no cycles, so it is a collection of paths, with at least one path $P$ of length at least 1. 
Choose an end vertex $v$ of $P$. Suppose first that $v\in J$.
Let $u$ be the vertex on $P$ that is adjacent to $v$ (so $u\in I$). Then in $I$, the token from $u$ can be moved to $v$, to obtain a new \indset\ $I'$.
In the case that $\meas(I,J)=1$ (the induction base), $|I\bs J|=1$, so $I'=J$ and we exhibited a \tsseq\ of length 1 between $I$ and $J$, which proves the claim.
Otherwise, note that $G[I'\Delta J]$ again contains no cycles, so by induction, there exists a \tsseq\ from $I'$ to $J$ of length at most $2(|I'\bs J|-1)\cdot \diam(G) + 2\md(I',J)\le$ $2(|I\bs J|-2)\cdot \diam(G) + 2\diam(G)=$ $2(|I\bs J|-1)\cdot \diam(G)$. 

Since $I'$ was obtained from $I$ using one token slide, we conclude that there exists a \tsseq\ from $I$ to $J$ of length at most $2(|I\bs J|-1)\cdot \diam(G)+1\le \meas(I,J)$, which proves the claim. If the chosen end vertex $v$ of the path $P$ is in $I$, then from $J$ we obtain $J'$ by sliding the adjacent token to $v$, and the statement can be proved analogously.

Now suppose that $\md(I,J)\ge 2$, let $d=\md(I,J)$. 
Choose $u\in I\bs J$ and $v\in J\bs I$ such that $\dist_G(u,v)=d$, and let $P=v_0,\ldots,v_d$ be a shortest path between $v_0=u$ and $v_d=v$.
We intend to slide the token from $u$ to $v$ along the path $P$. Define $I_i = I - u + v_i$ for $i=0,\dots,d$. 
If these are all \indset s, then they form a \tsseq\ of length $d$ from $I_0=I$ to a set $I_d$ that satisfies $|I_d\bs J|<|I\bs J|$. Then we can prove the statement by applying the induction assumption to $I_d$ and $J$, analogously to before.

Otherwise, let $i$ be the maximum index such that $I_i$ is not an \indset, and
let $x\in I-u$  be a token adjacent to $v_i$. (Informally: we choose a token $x$ adjacent to or on $P$, as close as possible to $v$. If $x$ lies on $P$, then this implies $x=v_{i-1}$.)
Note that $i<d$: Otherwise either $J$ is not an independent set (if $x\in J$) or $d=1$ (if $x\not\in J$), both contradictions. In addition, $i\ge 2$ holds: $i\ge 1$ is obvious, and if $i=2$ there there would be a $v_1$-claw with leaves $u$, $x$ and $v_2$.

We first argue that it is possible to slide the token from $x$ to $v$, along the path $x,v_i,v_{i+1},\ldots,v_d$.
By choice of $i$, there is no vertex in $I$ adjacent to $v_j$ for $j>i$. If there is a vertex $y\in I-x$ that is also adjacent to $v_i$, then $G$ contains a $v_i$-claw with leaves $x,y,v_{i+1}$, a contradiction. This shows that $I$ can be reconfigured to $I'=I-x+v$, using $d-i+1$ moves. 

It remains to show that we may apply the induction assumption to $I'$ and $J$. Clearly, $G[I'\Delta J]$ again contains no cycles.
Since $P$ is a shortest path, and $i\ge 2$, 
it holds that $\dist_G(x,v)\le d-i+1<d=\dist_G(u,v)$. 
So by choice of $u$ and $v$, it follows that $x\in J\cap I$. Therefore, $|I'\bs J|=|I\bs J|$. However, the minimum distance $\md(I',J)$ is now at most $d(u,x)$.
If $x\in V(P)$, then $x=v_{i-1}$ and $\dist_G(u,x)\leq i-1$; otherwise $x$ is adjacent to $v_{i-1}$ (since there is no $v_i$-claw with leaves $v_{i-1}$, $v_{i+1}$ and $x$), so $\dist_G(u,x)\le i$. Since $i<d=\md(I,J)$, we conclude that $\meas(I',J)<\meas(I,J)$, and thus we may apply the induction assumption to $I'$ and $J$. Combining this with the fact that we have a \tsseq\ from $I$ to $I'$ of length $d-i+1$, and that $d+i+1\le 2d=2\md(I,J)$, we conclude that there exists a \tsseq\ from $I$ to $J$ of length at most $d-i+1  + 2(|I'\bs J|-1)\cdot \diam(G) + 2\md(I',J)\le$ $d-i+1  + 2(|I\bs J|-1)\cdot \diam(G) + 2i\le$ $2(|I\bs J|-1)\cdot \diam(G) + 2\md(I,J)$. 
This concludes the proof of the induction step.
\QED

\section{Details for Section~\ref{sec:nonmaximum}}

First we show that in claw-free graphs, our definition of $I$-augmenting paths is equivalent with the definition used in the setting of finding maximum \indset s. The usual definition, as used e.g.\ in~\cite{schrijver2003combinatorial}, is given in the next proposition. 

\begin{propo}
\label{propo:AugmPathDefEquiv}
Let $I$ be an \indset\ in a claw-free graph $G$. 
An $I$-alternating walk $W=w_0,v_1,w_1,\ldots,v_k,w_k$ 
is an $I$-augmenting path if and only if 
$w_0,w_k\not\in I$ and $I'=I\bs \{v_1,\ldots,v_k\}\cup \{w_0,\ldots,w_k\}$ is an \indset. 
\end{propo}

\PF
Suppose $W$ is an $I$-augmenting path. Then clearly $w_0,w_k\not\in I$ and $v_i\in I$ for $i=0,\dots,k$. The vertices $w_0$ and $w_k$ have no neighbors in $I\bs \{v_1,\ldots,v_k\}$ since they are free. If a vertex $w_i$ with $1\le i\le k-1$ has a neighbor $x\in I\bs \{v_1,\ldots,v_k\}$, then $G$ contains a $w_i$-claw with leaves $v_{i},v_{i+1},x$, a contradiction. Two vertices $w_i$ and $w_j$ are not adjacent since $W$ is chordless. Hence $I'$ is an \indset\ again, which proves one direction of the statement.

Now suppose $I'$ is an \indset\ and $w_0,w_k\not\in I$. We prove that $W$ is an $I$-augmenting path. $W$ is chordless, otherwise there would be an edge $v_iw_j$ with $j\not\in \{i-1,i\}$ -- but then $G$ contains a $v_i$-claw with leaves $w_j,w_{i-1},w_i$, a contradiction. Therefore $w_0$ and $w_k$ are free with respect to $I$, so $W$ is an $I$-augmenting path.
\QED

We now give a detailed proof of Lemma~\ref{lem:non_maximum_NEW}. 
The proof is split up into three steps.

\begin{propo}
\label{propo:JustCyclesImpliesBothNondominating}
Let $A$ and $B$ be two \indset s in a claw-free graph $G$, such that $G[A\Delta B]$ is a collection of cycles. Then for any vertex $v\in V(G)$: if $N[v]\cap A=\emptyset$ then $N[v]\cap B=\emptyset$.
\end{propo}
\PF
Choose a vertex $v$ with $N[v]\cap A=\emptyset$, and suppose to the contrary that there exists a vertex $w\in N[v]\cap B$. 
Then $w\in B\bs A$, so $w$ is part of a cycle $C$ in $G[A\Delta B]$. Let $x$ and $y$ be the neighbors of $w$ on the cycle, so $\{x,y\}\subseteq A$. Then neither $x$ nor $y$ is adjacent to $v$, so there exists a $w$-claw with leaves $v,x,y$, a contradiction.
\QED

\begin{propo}
\label{propo:NonDomResolvingCycles}
Let $I$ and $J$ be \indset s in a claw-free graph $G$ such that $I$ is not a dominating set and $G[I\Delta J]$ is a collection of cycles. Then $I\tjr J$. 
\end{propo}
\PF
The proof is by induction over the number of cycles in $G[I\Delta J]$. If there are no cycles, then $I=J$ so $I\tjr J$ trivially holds. 

Now consider a cycle $C$ in $G[I\Delta J]$. Let $v$ be a vertex with $N[v]\cap I=\emptyset$ (which exists since $I$ is not dominating), and choose a vertex $u\in V(C)\cap I$. Then $I'=I+v-u$ is again an \indset, and clearly, $I\tjr I'$. 
Next, let $I''=I\Delta V(C)$, so $G[I\Delta I'']$ consists only of the cycle $C$.  Proposition~\ref{propo:JustCyclesImpliesBothNondominating} shows that $N[v]\cap I'=\emptyset$, so $G[I'\Delta I'']$ contains no cycles (it consists of one odd length path and one isolated vertex). Then by Theorem~\ref{thm:KMM}, $I'\tjr I''$, and thus $I\tjr I''$. Now $G[I''\Delta J]$ is again a collection of cycles, but contains exactly one cycle fewer than $G[I\Delta J]$ (namely $C$), so by induction we may conclude that $I''\tjr J$. Together, this shows that $I\tjr J$.
\QED

\PFof {\bf Lemma~\ref{lem:non_maximum_NEW}:}
First consider the case that $I$ is not a dominating set.
Let $H$ be the subgraph of $G[I\Delta J]$ that consists of all cycle components. (So possibly $H$ is the empty graph.) Let $I'=I\Delta V(H)$. So $G[I'\Delta J]$ contains no cycles, and $G[I\Delta I']=H$. Then by Theorem~\ref{thm:KMM}, $I'\tjr J$, and by Proposition~\ref{propo:NonDomResolvingCycles}, $I\tjr I'$. Together this shows that $I\tjr J$.

Otherwise, Proposition~\ref{propo:NotMaxImpliesAugmentable} shows that there exists an $I$-augmenting path $P$. Write $P=u_0,v_1,u_1,v_2,\ldots,v_k,u_k$, with $v_i\in I$ for all $i$. 
Then $I'=I\bs \{v_1,\ldots,v_k\}\cup \{u_1,\ldots,u_k\}$ is again an \indset\ with $|I'|=|I|$, and $G[I\Delta I']$ consists of a single (even length) path. So by Theorem~\ref{thm:KMM}, $I\tjr I'$.
Since $u_0$ is a free vertex for $I$, it is not dominated by $I'$. We conclude that $I'\tjr J$, and thus $I\tjr J$. 
\QED

\section{Proof Details for Lemmas~\ref{lem:resolvable_implies_turnable} and~\ref{lem:externally_or_internally}}

For our detailed proofs of the statements from Section~\ref{sec:resolving}, it is useful to first characterize the neighborhood of $I$-bad cycles using some simple observations (Proposition~\ref{propo:BadCycleNbhd}), and next characterize (shortest) resolving sequences (Lemma~\ref{lem:ShortResolvingSeqNEW}). 
For a vertex set $S\subseteq V(G)$, we denote $N(S)=\bigcup_{v\in S} N(v)$. 
(We will apply this only to \indset s $S$, so then $S\cap N(S)=\emptyset$.)

\begin{propo}
\label{propo:BadCycleNbhd}
Let $I$ be an \indset\ of a claw-free graph $G$, and let $C$ be an $I$-bad cycle with $I$-bipartition $[A,B]$. Then the following properties hold:
\begin{enumerate}[(a)]
 \item\label{pr:N3B_empty} 
 For all $i\ge 3$, $N_i(B)=\emptyset$.
 \item\label{pr:N2B_N0B}
 There are no edges between vertices in $N_2(B)$ and $N_0(B)$.
 \item\label{pr:NA_iff_NB}
 For any $v\in V(G)$: $v\in N(B) \bs A$ if and only if $v\in N(A)\bs B$. 
 \item\label{pr:NB_cap_I_eq_A}
 $N(B) \cap I = A$.  
\end{enumerate}
\end{propo}
\PF
\begin{enumerate}[(a)]
 \item This follows since $B$ is an \indset\ and $G$ is claw-free.
 \item Suppose to the contrary that $vw\in E(G)$ with $v\in N_2(B)$ and $w\in N_0(B)$. Let $N(v)\cap B=\{x,y\}$. Then $G$ contains a $v$-claw with leaves $v,x,y$, a contradiction.
 \item Suppose $v \in N(B)\bs A$ and $v\not\in N(A)\bs B$ (the symmetric case is analogous). Since $v\in N(B)$, we have $v\not\in B$, so $v\not\in N(A)$. Choose any vertex $x \in B\cap N(v)$. It has two neighbors $y,z\in A$, so $G$ contains an $x$-claw with leaves $v,y,z$, a contradiction.
 \item Suppose there is a token $v\in I\bs A$ in the neighborhood of $B$. By the previous claim, $v \in N(A)$, which contradicts that $I$ is an independent set.\QED
\end{enumerate}

{\em Shortest} resolving sequences have a very specific and useful structure, which is characterized in the following lemma. In particular, this lemma shows that a \tsseq\ (starting with $I$) resolves an $I$-bad cycle with $I$-bipartition $[A,B]$ as soon as the first token slides from  $N_2(B)$ to $N_1(B)$, but not earlier.
Let $C$ be an $I$-bad cycle with $I$-bipartition $[A,B]$.
We say that a sequence $I_0,\ldots,I_m$ with $I_0=I$ {\em contains a resolving sequence} for $C$ if for some $i\in \{0,\ldots,m\}$, $G[I_i\cup B]$ contains no cycle (so $I_0,\ldots,I_i$ is a resolving sequence, although $I_0,\ldots,I_m$ may not be one).

\begin{lem}
\label{lem:ShortResolvingSeqNEW}
Let $I$ be an \indset\ of a claw-free graph $G$, and let $C$ be an $I$-bad cycle with $I$-bipartition $[A,B]$. 
Let $S=I_0,\ldots,I_m$ be a \tsseq\ with $I_0=I$. 
Then the following properties hold:
\begin{enumerate}[(a)]
 \item\label{pr:a} 
 $S$ contains a resolving sequence for $C$ if and only if it contains a move $u\to v$ with $u\in N_2(B)$ and $v\in N_1(B)$.
 \item\label{pr:b} 
 For every index $i$ such that $I_0,\ldots,I_i$ contains no resolving sequence for $C$: 
 $I_i\subseteq N_2(B)\cup N_0(B)$, and 
 $I_i$ is obtained from $I_{i-1}$ by a move $u\to v$ with $N(u)\cap B=N(v)\cap B$.
\end{enumerate}
\end{lem}

\PF
Call an \indset\ $J$ {\em $B$-cyclic} if there is one cycle in $G[J\cup B]$ that contains all vertices of $B$. 
Since $J$ is an \indset\ and $G[J\cup B]$ has maximum degree 2, 
this implies that $G[J\cup B]$ consists of exactly one cycle and a number of isolated vertices. Furthermore, it implies that every $v\in B$ has exactly two neighbors in $J$, which in turn are in $N_2(B)$.

Now consider an \indset\ $I_{i}$ in the sequence $S$, that is obtained from a $B$-cyclic set $I_{i-1}$ using the move $u\to v$.  Since any vertex in $B$ has at most two neighbors in any \indset\ of $G$, and $I_{i-1}$ is $B$-cyclic, we deduce that $N(v)\cap B\subseteq N(u)\cap B$. Clearly, $I_i$ is again $B$-cyclic if and only if $N(v)\cap B=N(u)\cap B$. 
So if $I_i$ is not $B$-cyclic, then $|N(v)\cap B|\le 1$ and $|N(u)\cap B|\ge 1$. Since $I_{i-1}\subseteq N_0(B)\cup N_2(B)$, it follows that $u\in N_2(B)$, and since vertices in $N_2(B)$ have no neighbors in $N_0(B)$ (Proposition~\ref{propo:BadCycleNbhd}(\ref{pr:N2B_N0B})), it follows that $v\in N_1(B)$. In this case, we argue that $G[I_i\cup B]$ contains no cycle: If to the contrary $G[I_i\cup B]$ contains a cycle $C'$, then $C'$ does not contain $v$. So it would also be a cycle in $G[I_{i-1}\cup B]$, which contains neither $u$ nor its neighbors in $B$, contradicting that $I_{i-1}$ is $B$-cyclic.
Summarizing, we have shown that if $I_i$ is obtained from $I_{i-1}$ using the move $u\to v$ and $I_{i-1}$ is $B$-cyclic, then:
\begin{enumerate}[(i)]
 \item\label{pr:i}
 If $I_i$ is again $B$-cyclic, then $N(u)\cap B=N(v)\cap B$.
 \item\label{pr:ii}
 If $I_i$ is not $B$-cyclic, then $u\in N_2(B)$ and $v\in N_1(B)$, and $G[I_i\cup B]$ contains no cycles.
\end{enumerate}
We use this to prove the properties in the lemma statement. Note that $I_0=I$ is $B$-cyclic, so if $S$ contains a non-$B$-cyclic set, then the first such set $I_i$ has $i\ge 1$ and is preceded by a $B$-cyclic set $I_{i-1}$.

If $S$ contains a resolving sequence, then clearly it contains a non-$B$-cyclic set, so by considering the first non $B$-cyclic set $I_i$ and applying~(\ref{pr:ii}), we conclude that $S$ contains a move from $N_2(B)$ to $N_1(B)$. 
On the other hand, if $S$ contains such a move, then from~(\ref{pr:i}) it follows that $S$ contains a non-$B$-cyclic set, and therefore by~(\ref{pr:ii}), it contains a resolving sequence for $C$. This proves Property~(\ref{pr:a}). 

Property~(\ref{pr:ii}) implies that if a subsequence $I_0,\ldots,I_i$ contains no resolving sequence for $C$, then all these sets are $B$-cyclic, so Property~(\ref{pr:b}) follows from~(\ref{pr:i}).
\QED

Now we can prove Lemma~\ref{lem:resolvable_implies_turnable} and Lemma~\ref{lem:externally_or_internally} in detail.

\PFof {\bf Lemma~\ref{lem:resolvable_implies_turnable}:}
Denote $J=I\Delta V(C)$, and let $[A,B]$ be the $I$-bipartition of $C$.
Consider a {\em shortest} \tsseq\ $I_0,\ldots,I_m$ that resolves $C$. Suppose first that $m=1$, so $I_1$ is obtained from $I$ by a move $u\to v$ with $u\in N_2(B)$ and $v\in N_1(B)$ (Lemma~\ref{lem:ShortResolvingSeqNEW}(\ref{pr:a})). Since $G$ contains no $u$-claw, it follows that $N(v)\cap B\subset N(u)\cap B$, so we can label the vertices of $C$ $c_1,\ldots,c_{2n}$ in order along the cycle such that $u=c_1$ and $N(v)\cap B=\{c_2\}$. 
Then the following sequence of moves yields $J$, when starting with $I$:
$c_1\to v$, $c_{2n-1}\to c_{2n}$, $c_{2n-3}\to c_{2n-2},\ldots,c_3\to c_4,v\to c_2$.
Using the fact that $C$ is a chordless cycle and that $N(B)\cap I=A$
(Proposition~\ref{propo:BadCycleNbhd}(\ref{pr:NB_cap_I_eq_A})),
it is easily verified that every vertex set in the resulting sequence is an \indset, so $I\tsr J$.

Now suppose that $m\ge 2$. Let $I'=I_{m-1}$. Then there exists an $I'$-bad cycle $[A',B]$, since until this point in the \tsseq, tokens that started on $A$ (i.e. tokens on $N_2(B)$) only moved to vertices with exactly the same neighbors in $B$ (Lemma~\ref{lem:ShortResolvingSeqNEW}(\ref{pr:b})). From $I'$ we can obtain $J'=I'\Delta V(C)$ in the same way as shown the previous paragraph. 
It remains to show that from $J'$, $J$ can be obtained, by essentially reversing all moves outside the neighborhood of the cycle, while moving no tokens on B.
More precisely, for every $i\in \{0,\ldots,m-1\}$, define $I'_i=(I_i\bs N(B))\cup B$.
Note that $I'_0=J$ and that $I'_{m-1}=J'$. We argue that (after removing repetitions), $I'_{m-1},\ldots,I'_0$ yields a \tsseq\ from $J'$ to $J$:
By Lemma~\ref{lem:ShortResolvingSeqNEW}(\ref{pr:b}), for any $i<m-1$, if $I'_{i+1}\not=I'_i$, then $I'_{i+1}$ can be obtained from $I'_i$ by a move $u\to v$ where both $u$ and $v$ are part of $N_0(B)$. 
This way, it can be verified that for every $i$, $I'_i$ is an \indset, so $J'\tsr J$. Combining this with $I\tsr I'$ and $I'\tsr J'$ shows that $I\tsr J$.\QED

\PFof {\bf Lemma~\ref{lem:externally_or_internally}:}
Denote the $I$-bipartition of $C$ by $[A,B]$.
Consider a {\em shortest} \tsseq\ $S=I_0,I_1,\dots,I_m$ that resolves $C$. 
Let $u\to v$ be the last move of this sequence, so $u\in N_2(B)$ and $v\in N_1(B)$ (Lemma~\ref{lem:ShortResolvingSeqNEW}(\ref{pr:a})). By Proposition~\ref{propo:BadCycleNbhd}(\ref{pr:NA_iff_NB}), $|N(v)\cap A|\ge 1$, and clearly, $|N(v)\cap I|\le 2$. 
So one of the following cases applies to the neighborhood of $v$. 

\medskip
\noindent
{\bf Case 1:} $N(v)\cap I=\{x\}$ for some $x\in A$.\\
Then the move $x\to v$ yields an \indset\ again, and since $v\in N_1(B)$, it resolves $C$ (Lemma~\ref{lem:ShortResolvingSeqNEW}(\ref{pr:a})), so $C$ is both internally and externally resolvable.

\medskip
\noindent
{\bf Case 2:} $N(v)\cap I=\{x,y\}$ for some $x\in A$ and $y\not\in A$.\\
In this case, we omit all internal moves, to obtain an external \tsseq\ that resolves $C$. More precisely, for every $i$, define $I'_i=(I_i\bs N(B))\cup A$, and consider the sequence $I'_0,\ldots,I'_{m-1}$. For every $i$ such that $I'_i\not=I'_{i+1}$, it holds that $I'_{i+1}$ is obtained from $I'_i$ by a move $u_i\to v_i$ where both $u_i$ and $v_i$ are in $N_0(B)$ (Lemma~\ref{lem:ShortResolvingSeqNEW}(\ref{pr:b})). Since there are no edges between $N_0(B)$ and $A\subseteq N_2(B)$ (Proposition~\ref{propo:BadCycleNbhd}(\ref{pr:N2B_N0B})), every $I'_i$ is an \indset, and thus this is a \tsseq. 
Now $x\in I'_{m-1}$ because $x\in A$, and $I_{m-1}\cap N(v)=\{u\}\subseteq N_2(B)$, so $N(v)\cap I'_{m-1}=\{x\}$. Therefore, from $I'_{m-1}$, the move $x\to v$ can be made, to resolve $C$ (Lemma~\ref{lem:ShortResolvingSeqNEW}(\ref{pr:a})). This shows that $C$ is externally resolvable.

\medskip
\noindent
{\bf Case 3:} $N(v)\cap I=\{x,y\}$ for some $x,y\in A$.\\
In this case, we omit all external moves, to obtain an internal \tsseq\ that resolves $C$.
More precisely, for every $i$, define $I'_i=(I\bs N(B))\cup (I_i\cap N(B))$, and consider the sequence $I'_0,\ldots,I'_{m-1}$. For every $i<m-1$ such that $I'_i\not=I'_{i+1}$, it holds that $I'_{i+1}$ is obtained from $I'_i$ by a move $u_i\to v_i$ where both $u_i$ and $v_i$ are in $N_2(B)$ (Lemma~\ref{lem:ShortResolvingSeqNEW}(\ref{pr:b})). Since there are no edges between $N_2(B)$ and $(I\bs N(B))\subseteq N_0(B)$ (Proposition~\ref{propo:BadCycleNbhd}(\ref{pr:N2B_N0B})), every $I'_i$ is an \indset, and thus this is a \tsseq. 
Since $I_{m-1}\cap N(v)=\{u\}$ and $N(v)\cap I\subseteq N(B)$, it also holds that $I'_{m-1}\cap N(v)=\{u\}$, so from $I'_{m-1}$, the move $u\to v$ can be made, to resolve $C$ (Lemma~\ref{lem:ShortResolvingSeqNEW}(\ref{pr:a})). This shows that $C$ is internally resolvable.
\QED

\section{The Proof of Theorem~\ref{thm:ext_res_augm_path}}

\begin{figure}
\centering
\scalebox{1}{$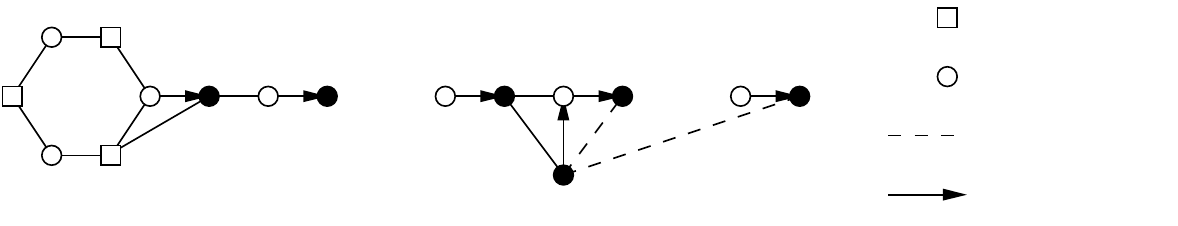$}
\caption{An illustration of the proof of Lemma~\ref{lem:tokens_move_once}.}
\label{fig:MoveOncePf}
\end{figure}

We prove in this section that to verify whether an $I$-bad cycle $C$ is externally resolvable it suffices to search for a certain type of $I$-augmenting paths in $G-V(C)$. The key observation is that in a shortest \tsseq\ that externally resolves $C$, no token moves more than once, provided that $I$ is a maximum \indset\ (Lemma~\ref{lem:tokens_move_once} below). In that case, the token moves easily yield a set of $I$-alternating paths, as shown in the next lemma.
Note that Lemma~\ref{lem:tokens_move_once} may fail if we drop the assumption that $I$ is a maximum \indset\ 
(for example, consider the graph from Figure~\ref{fig:MoveOncePf} with dashed edges removed, and let $I$ contain vertex $x$ together with all round white vertices, except for $y$. We note that $I$ is then maximal, but not maximum). 
To be precise, we say that in a \tsseq\ $I_0,\ldots,I_n$, {\em every token moves at most once} if for all $i,j\in \{0,\ldots,n-1\}$ with $i\le j$:
$v\in I_{j}\bs I_{j+1}$ implies that $v\in I_i$.

\begin{propo}
\label{propo:moving_once_paths}
Let $I_0,\dots,I_m$ be a \tsseq\ in a claw-free graph $G$ in which every token moves at most once. 
Then every component of $G[I_0\Delta I_m]$ is a path of odd length.
\end{propo}

\PF
The proof is by induction on $m$. 
Let $I_m$ be obtained from $I_{m-1}$ by the move $x\to y$.
By induction, $G'=G[I_0\Delta I_{m-1}]$ is a set of paths of odd length.
Since every token moves at most once, $G'$ contains neither $x$ nor $y$.
By claw-freeness, $G[I_0\Delta I_m]$ has maximum degree two, and since it is obtained from $G'$ by adding two adjacent vertices (plus incident edges), no even length path can be introduced. So it now suffices to show that in $G[I_0\Delta I_m]$, there is no cycle containing $x$ and $y$. 
The vertex $x$ is part of both $I_0$ and $I_{m-1}$, so it has no neighbors in $I_0\Delta I_{m-1}$. Therefore it has degree 1 in $G[I_0\Delta I_m]$, which shows that it is not part of a cycle, and thus $G[I_0\Delta I_m]$ is a collection of odd length paths again.\QED

We will often use the following simple proposition.

\begin{propo}
\label{propo:odd_paths_are_good}
Let $I$ and $J$ be two \indset s in a graph $G$. 
If every component in $G[I\Delta J]$ is an odd length path, and $|I\bs J|=p$, then there exists a \tsseq\ from $I$ to $J$ of length $p$.
\end{propo}

\PF
We prove the statement by induction on $|I\bs J|$. The case $I=J$ is trivial, so now assume that there exists at least one odd length path component $P$ in $G[I\Delta J]$. 
Then $P$ has an end vertex $v\in J\bs I$ with neighbor $u\in I\bs J$, but with no other neighbors in $I$. So from $I$, we can make the move $u\to v$, which yields $I'$, such that every component of $G[I'\Delta J]$ is again an odd length path, and $|I'\bs J|=p-1$. The statement now follows by induction.
\QED

\begin{lem}
\label{lem:tokens_move_once}
Let $I$ be a maximum independent set in a claw-free graph $G$, and let $C$ be an externally resolvable $I$-bad cycle. 
Then in any shortest external resolving sequence for $C$, every token moves at most once.
\end{lem}
\PF
The following proof is illustrated in Figure~\ref{fig:MoveOncePf}.
Let $[A,B]$ be the $I$-bipartition of $C$.
Let $S=I_0,\ldots,I_m$ be a shortest external resolving sequence for $C$. By definition of {\em external resolving sequence}, all moves but the last one are between vertices in $N_0(B)$, and the last move of the sequence is $u\to v$ for some $u\in N_2(B)$ and $v\in N_1(B)$ (Lemma~\ref{lem:ShortResolvingSeqNEW}(\ref{pr:a})). We prove the statement by induction on $m$. If $m=1$ then obviously, no token moves twice.

Now suppose that $m\ge 2$. Then $A\subseteq I_1$, so $C$ is also an $I_1$-bad cycle with $I_1$-bipartition $[A,B]$, and $I_1,\ldots,I_m$ is a shortest external resolving sequence for $C$ with respect to $I_1$. By induction, no token moves twice in this sequence. So every component of $G[I_1\Delta I_m]$ is an odd length path (Proposition~\ref{propo:moving_once_paths}), which is clearly both $I_1$-alternating and $I_m$-alternating.
Let $P=v_1,u_1,v_2,u_2,\ldots,v_k,u_k$ be the path in $G[I_1\Delta I_m]$ that contains $u$ and $v$, labeled such that $u_i\in I_1$ and $v_i\in I_m$ for every $i$. 
So $u=u_{k'}$ and $v=v_{k'}$ for some index ${k'}$. Then it is easily verified that from $I_1$, we can make the sequence of moves $u_1\to v_1,\ldots,u_{k'}\to v_{k'}$, maintaining an \indset\ throughout.
This \tsseq\ resolves $C$ using ${k'}$ moves (Lemma~\ref{lem:ShortResolvingSeqNEW}(\ref{pr:a})).
Since the \tsseq\ $I_1,\ldots,I_m$ is a shortest \tsseq\ for $I_1$ that resolves $C$, we conclude that these are all moves from the sequence. So $k=k'$, and the path $P$ is the only component in $G[I_1\Delta I_m]$, and thus $m=k+1$.

If in the entire sequence $I_0,\ldots,I_m$ no token moves twice, then there is nothing to prove, so now assume that at least one token moves twice. Since no token moves twice in the subsequence $I_1,\ldots,I_m$, the first move from $I_0$ to $I_1$ is $x\to y$, and later in the sequence, a move $y\to z$ occurs. Then $y$ and $z$ lie on the path $P$, so $y=u_j$ and $z=v_j$ for some index $j$.
We start with a few simple observations:
\begin{enumerate}
 \item $j<k$. 
 
 If to the contrary $j=k$, then $y=u$ and $z=v$, so $y\in N_2(B)$. But then $x\in N_2(B)$ (Lemma~\ref{lem:ShortResolvingSeqNEW}(\ref{pr:b})), which contradicts that $I_0,\ldots,I_m$ is an {\em external} resolving sequence.
 
 \item $N(x)\cap V(P)\subseteq \{v_1,v_j,u_j,v_{j+1}\}$.
 
 This holds because $x$ cannot be adjacent to a vertex $u_{j'}$ with $j'\not=j$, since these vertices are both part of the \indset\ $I_0$. Furthermore, any vertex $v_{j'}$ with $j'\not\in \{1,j,j+1\}$ has neighbors $u_{j'}$ and $u_{j'-1}$, which are both in $I_0$, so an edge $xv_{j'}$ would yield a $v_{j'}$-claw.

 \item $xv_{j+1}\in E(G)$. 
 
 Assume to the contrary that $xv_{j+1}\not\in E(G)$. 
 We can then argue that $J:=(I_0\bs \{u_{j+1},\ldots,u_k\})\cup \{v_{j+1},\ldots,v_k\}$ is also an \indset: $P$ is chordless, so none of the {\em added vertices} $\{v_{j+1},\ldots,v_k\}$ are adjacent to vertices in $\{u_1,\ldots,u_{j-1}\}$. By the previous observation and the assumption $xv_{j+1}\not\in E(G)$, $x$ is also not adjacent to any of the added vertices. Finally, considering $I_m$, which contains the added vertices and the vertices $I_0\bs V(P)$, we conclude that the added vertices are not adjacent to vertices of $I_0\bs V(P)$.
 Note that $G[I_0\Delta J]$ consists of a single odd path on $2(k-j)$ vertices, so there exists a (shorter) \tsseq\ of length $k-j$ which resolves $C$ (Proposition~\ref{propo:odd_paths_are_good}), a contradiction.
\end{enumerate}

To complete the proof we consider four cases.

\medskip
\noindent
{\bf Case 1:} $v_{j}\notin N(x)$ and $v_1\notin N(x)$.\\
If $j=1$, then $v_1=v_j$ has no neighbors in $I_0$. Otherwise, both $v_1$ and $v_j$ are free vertices with respect to $I_0$ (their only $I_0$-neighbors are $u_1$ and $u_{j-1}$, respectively), so $v_1,u_1,\ldots,u_{j-1},v_j$ is an $I_0$-augmenting path. In both cases, this contradicts that $I=I_0$ is a maximum \indset.

\medskip
\noindent
{\bf Case 2:} $v_{j}\in N(x)$ and $v_1\notin N(x)$.\\
Then the previous observations show that in $G[I_0\Delta I_m]$, we have the following odd path component: 
\[v_1,u_1,\ldots,v_j,x,v_{j+1},u_{j+1},\ldots,v_k,u_k,
\]
containing $2k$ vertices. Hence a \tsseq\ of shorter length $k<m$ that resolves $C$ is possible (Proposition~\ref{propo:odd_paths_are_good}), a contradiction.

\medskip
\noindent
{\bf Case 3:} $v_{j}\notin N(x)$ and $v_1\in N(x)$.\\
Then the previous observations show that in $G[I_0\Delta I_m]$, we have the following odd path component: 
\[v_j,u_{j-1},v_{j-1},\ldots,u_1,v_1,x,v_{j+1},u_{j+1},\ldots,v_k,u_k,\]
containing $2k$ vertices. Hence a \tsseq\ of shorter length $k<m$ that resolves $C$ is possible (Proposition~\ref{propo:odd_paths_are_good}), a contradiction.

\medskip
\noindent
{\bf Case 4:} $v_{j}\in N(x)$ and $v_1\in N(x)$.\\
If $j=1$, then $y=u_j=u_1$ and the first two moves $x\to y, u_1\to v_1$ can be replaced by one move $x\to v_1$, giving a shorter \tsseq, a contradiction. Otherwise, $x$ has three neighbors $v_1$, $v_j$ and $v_{j+1}$ in an independent set $I_m$, contradicting claw-freeness.

\medskip
\noindent
We have obtained a contradiction in every case, so we conclude that in a shortest external \tsseq\ of length $m$, no token moves twice. This concludes the inductive step of the proof, and the statement follows by induction.
\QED

We can now combine Proposition~\ref{propo:moving_once_paths} and Lemma~\ref{lem:tokens_move_once} to prove Theorem~\ref{thm:ext_res_augm_path}.

\PFof {\bf Theorem~\ref{thm:ext_res_augm_path}:}
Denote $G'=G-A-B$ and $I'=I\bs A$, so $I'$ is an \indset\ of $G'$.

Suppose first that $G'$ contains such an $I'$-augmenting path $P=u_0$,$v_0$,$u_1$,$v_1$\\,$\dots$,$v_{k-1}$,$u_k$ with $u_0\in N_0(B)$ and $u_k\in N_1(B)$. (So $k\ge 1$.) We prove that then $C$ is externally resolvable. 
Since $u_0\not\in N(B)$ and $u_0\not\in B$, we observe that $u_0\not\in N(A)$ (Proposition~\ref{propo:BadCycleNbhd}(\ref{pr:NA_iff_NB})).
Secondly, we argue that $|N(u_k)\cap A|=1$: by Proposition~\ref{propo:BadCycleNbhd}(\ref{pr:NA_iff_NB}), $u_k$ has at least one neighbor $w$ in $A$. Since $u_k$ is also adjacent to $v_{k-1}\in I\bs A$ and there is no $u_k$-claw, $w$ is its only neighbor in $A$. All other vertices $u_j$ with $1\le j\le k-1$ are adjacent to $v_{j-1}$ and $v_j$, which are both in $I\bs A$, so they have no other neighbors in $I$, in particular not in $A$. 
Since $u_0$ is free with respect to $I'$, it has no neighbor in $I$ other than $v_0$.
Since $P$ is also chordless, it follows that we can apply the moves $v_0\to u_0$,\ldots,$v_{k-1}\to u_{k-1}$ to $I$ while maintaining an \indset, which yields $J$. We have that $N(u_k)\cap J=\{w\}$, so the move $w\to u_k$ (with $w\in A$ and $u_k\in N_1(B)$) is subsequently possible, and resolves the cycle $C$ (Lemma~\ref{lem:ShortResolvingSeqNEW}(\ref{pr:a})), and thus $C$ is externally resolvable.

\medskip
We now prove the other direction. Suppose that $C$ is externally resolvable, and consider a shortest external resolving sequence $S=I_0,\ldots,I_m$ (with $I_0=I$).  By definition of {\em external resolving sequence} and Lemma~\ref{lem:ShortResolvingSeqNEW}(\ref{pr:a}), the last move is  $u\to v$ for some $u\in N_2(B)$ and $v\in N_1(B)$, and every other move is between two vertices in $N_0(B)$. By Lemma~\ref{lem:tokens_move_once}, every token moves at most once in $S$. So $G[I_0\Delta I_m]$ is a set of odd paths (Proposition~\ref{propo:moving_once_paths}), in which all vertices except $u$ and $v$ are in $N_0(B)$. Clearly these paths are all both $I_0$-alternating and $I_m$-alternating.
Consider the path $P$ that contains $u$ and $v$ and denote $P=v_0,u_0,v_1,u_1,\ldots,v_k,u_k$, with $v_i\in I_m$ and $u_i\in I_0$ for all $i$. 
So for every $i$, $u_i\to v_i$ is a move in the \tsseq\ $S$, and $u_k\to v_k$ must be the last of these moves on $P$, since an \indset\ should be maintained throughout. So $u=u_k$ and $v=v_k$.
We now argue that $P'=v_0,u_0,v_1,u_1,\ldots,u_{k-1},v_k$ is the desired $I'$-augmenting path in $G'$: clearly the path is $I'$-alternating and chordless. The vertex $v_0$ is not adjacent to any $I$-vertex $x$ other than $u_0$, because otherwise $I_m$ is not an \indset\ (if $x\in I_m$), or $P$ is not a component of $G[I_0\Delta I_m]$ (if a move $x\to y$ occurs in $S$). So $v_0$ is a free vertex with respect to both $I$ and $I'$.
The vertex $v_k=v$ is adjacent to both $u_{k-1}$ and $u_k=u$ (in $G$), so it is not adjacent to any other vertex from $I$. Therefore, in $G'$, it is a free vertex with respect to $I'$ (which does not contain $u$). This shows that $P'$ is an $I'$-augmenting path for $G'$, between vertices in $N_0(B)$ and $N_1(B)$.
\QED

\section{The Proof of Theorem~\ref{thm:internally_iff_copath}}

Lemma~\ref{lem:internally_implies_copath} proves the forward direction of Theorem~\ref{thm:internally_iff_copath}, and subsequently, Lemma~\ref{lem:copath_implies_internally} proves the backward direction.

\begin{lem}
\label{lem:internally_implies_copath}
Let $I$ be an independent set in a claw-free graph $G$, and $C=c_0,c_1,\dots,c_{2n-1},c_{0}$ be an $I$-bad cycle with $n\ge 3$ and $c_0\in I$, and $I$-bipartition $[A,B]$. 
If $C$ is internally resolvable then $D(G,C)$ or $D(G,C^{rev})$ contains a directed path from a vertex $b\in N_1(B)$ with $N(b)\cap I\subseteq A$ to a vertex in $A$.
\end{lem}

\PF Let $I_0,\dots,I_m$ be a shortest internal resolving \tsseq\ for $C$, so $I_0=I$.
By definition of {\em internally resolvable} and Lemma~\ref{lem:ShortResolvingSeqNEW}(\ref{pr:a}), the last move is is from $N_2(B)$ to a vertex $b\in N_1(B)$, and all other moves $u\to v$ satisfy $u,v\in N_2(B)$ and $N(u)\cap B = N(v)\cap B$.
We shall prove that $A$ is reachable from $b$ by a directed path in $D(G,C)$ or $D(G,C^{rev})$.

For every $i\in \{0,\ldots,n-1\}$, $I_0$ contains exactly one vertex of $L_i$ (namely $c_{2i}$), and this accounts for all vertices in $I_0\cap N_2(B)$. Since $N(u)\cap B = N(v)\cap B$ holds for every move $u\to v$, this property is maintained for every $I_j$.
So for all $j\in\{0,\dots, m-1\}$ and $i\in\{0,\ldots,n-1\}$, we may denote by $v^i_j$ the unique vertex in $I_j \cap L_i$.

W.l.o.g.\ assume $N(b)\cap B=\{c_1\}$.
Since a token is moved to $b$ in the last move, $b$ is free in $I_{m-1}$, so it cannot be adjacent to both $v^{0}_{m-1}$ and $v^{1}_{m-1}$. Assume w.l.o.g.\ that it is nonadjacent to $v^1_{m-1}$ (otherwise reverse the order of vertices of $C$, such that the remainder of the proof applies to $D(G,C^{rev})$ instead of $D(G,C)$). Then by definition, $D(G,C)$ contains an arc $(b,v^1_{m-1})$. 
For all $j$, denote $I'_j=I_j\cap N(B)$. So we conclude that at least one vertex of $I'_{m-1}$ is reachable from $b$ in $D(G,C)$.

For every $j=0,\dots,m-1$ and every $i$, $D(G,C)$ contains an arc from $v^i_j$ to $v^{(i+1)\bmod n}_j$, because these vertices are both in $I_j$ and are therefore nonadjacent in $G$. Therefore if for some $j=0\dots m-1$, at least one vertex in $I'_j=\{v^0_j,\dots,v^{n-1}_j\}$ is reachable from $b$ in $D(G,C)$, then all vertices of $I'_j$ are reachable. But $I'_{j-1}\cap I'_j\not=\emptyset$ (they share in fact $n-1$ vertices), so by a simple induction proof it follows that every vertex of every $I'_j$ is reachable from $b$ in $D(G,C)$. 
In particular, this shows that there is a directed path from $b$ to $I'_0=A$ in $D(G,C)$.

It remains to show that $b$ has no neighbors in $I\bs A$. Suppose that some vertex $x\in I\bs A$ is adjacent to $b$. Since $x\notin N_2(B)$ (Proposition~\ref{propo:BadCycleNbhd}(\ref{pr:NB_cap_I_eq_A})), the token on $x$ is never moved in the \tsseq, so $x\in I_m$. But its neighbor $b$ is also part of the independent set $I_m$, a contradiction.
\QED

\begin{lem}
\label{lem:copath_implies_internally}
Let $I$ be an independent set in a claw-free graph $G$, and $C=c_0,c_1,\dots,c_{2n-1},c_{0}$ be an $I$-bad cycle with $n\ge 4$ and $c_0\in I$, and $I$-bipartition $[A,B]$. 
If $D(G,C)$ or $D(G,C^{rev})$ contains a directed path from a vertex $b\in N_1(B)$ with $N(b)\cap I\subseteq A$ to a vertex in $A$, then $C$ is internally resolvable.
\end{lem}
\PF
W.l.o.g.\ we may assume that $D(G,C)$ contains such a path, and that $N(b)\cap B=\{c_1\}$.
Let $P=u_0,\dots,u_m$ be a shortest path in $D(G,C)$ from $u_0=b$ to a vertex $u_m\in A$.
Throughout this proof, we take layer indices modulo $n$ and cycle indices modulo $2n$, so $L_i$ denotes $L_{i \bmod n}$ and $c_i$ denotes $c_{i\bmod 2n}$.
By definition of $D(G,C)$, any arc $(u_0,x)$ must have $x\in L_1$. Thus $u_1\in L_1$ and similarly, $u_j\in L_j$ follows inductively for all $j=1,\dots,m$.
The vertex $u_m$ is the first vertex of $P$ in $A$, with $u_m\in L_m$ so $u_m=c_{2m}$.
For all $j$, $D(G,C)$ contains an arc from $c_{2j}$ to $c_{2j+2}$, so we can extend $P$ to a directed path $P'=u_0,\ldots,u_{m+n-1}$ by defining $u_{m+j}=c_{2(m+j)}$ for $j=1\dots n-1$. The following properties hold for $P'$, and will be used often in the remainder of the proof:
\begin{align*} 
 & N(u_0) \cap B = \{c_1\},\\
 & N(u_j) \cap B = \{c_{2j-1}, c_{2j+1}\} \mbox{ for } j=1,\dots,m+n-1.
\end{align*}

The idea is to reconfigure $A$ to subsequent infixes of $P'$.  More precisely, for $j=0,\dots,m$, define
\[I_j = \{u_{m-j},u_{m-j+1},\dots,u_{m-j+n-1}\} \cup (I\bs A),\]
and consider the sequence $I_0,\ldots,I_m$. This sequence starts with $I_0 = I$ (since $A=\{u_m,\ldots,u_{m+n-1}\}$).
Since $I_j=I_{j-1}+u_{m-j}-u_{m-j+n}$ for $j=1,\dots,m$, the consecutive steps correspond to replacing $u_{m-j+n}$ by $u_{m-j}$. 
We will now show that for every $j$, $I_j$ is an \indset, and that $u_{m-j+n}u_{m-j}\in E(G)$ (so $u_{m-j+n}\to u_{m-j}$ is a valid move), which shows that $I_0,\ldots,I_m$ is a \tsseq. Considering the last move, it then follows that this sequence resolves $C$ (Lemma~\ref{lem:ShortResolvingSeqNEW}(\ref{pr:a})), and is in fact an internal resolving sequence.
Summarizing, to prove the lemma, it now suffices to show that:
\begin{enumerate}[(a)]
 \item\label{it:move}
 $u_{m-j}u_{m-j+n}\in E(G)$ for all $j\in \{1,\ldots,m\}$, and
 \item\label{it:ind}
 $u_ju_{j'}\not\in E(G)$, for all $j,j'\in\{0,\dots,m+n-1\}$ with $1\leq j'-j \leq n-1$.
\end{enumerate}
Indeed, the second condition ensures that each $I_j$ is an independent set: $u_0=b$ has no neighbors in $I\bs A$ by assumption, and the vertices $u_j$ for $j\ge 1$ have no neighbors in $I\bs A\subseteq N_0(B)$ because there are no edges between $N_0(B)$ and $N_2(B)$ (Proposition~\ref{propo:BadCycleNbhd}(\ref{pr:N2B_N0B})).

\medskip 
To prove the above statements, we will prove a few claims, that are marked with Greek letters for later reference.

{\bf Claim $\alpha$}:\quad $u_{j}u_{j+1} \notin E(G)$ for all $j$.

This claim follows directly from the definition of $D(G,C)$ and the fact that these are consecutive path vertices.

{\bf Claim $\beta$}:\quad $u_{j}u_{j'} \notin E(G)$ for $2\le j'-j\le n-2$.

If $j>0$ then $u_j$ and $u_{j'}$ belong to $L_j$ and $L_{j'}$, respectively. So their neighbors in $B$ are exactly $c_{2j-1},c_{2j+1}$ and $c_{2j'-1},c_{2j'+1}$, respectively. By choice of $j$ and $j'$, 
these are four different vertices. 
Thus if $u_ju_{j'}\in E(G)$, then there would be a $u_{j'}$-claw with leaves $u_j,c_{2j'-1},c_{2j'+1}$, a contradiction. 
If $j=0$, then the proof is analogous, except that $u_j$ has exactly one neighbor in $B$, namely $c_1=c_{2j+1}$. This concludes the proof of Claim~$\beta$.

Together, Claims~$\alpha$ and $\beta$ prove statement~(\ref{it:ind}) above for all cases except $j'-j=n-1$. So we conclude that it now remains to show that:

{\bf Claim $\gamma$}:\quad $u_ju_{j+n}\in E(G)$ for all $j\in \{0,\dots,m-1\}$ and

{\bf Claim $\delta$}:\quad $u_ju_{j+n-1}\not\in E(G)$ for all $j\in\{0,\dots,m\}$.

\noindent We prove these claims by induction on $j$.

\smallskip 
$\delta(0)$:\quad $u_0u_{n-1}\not\in E(G)$, for otherwise there would be a $u_{n-1}$-claw with leaves $u_0,c_{2n-3},c_{2n-1}$ (recall that $N(u_0)\cap B=c_1$, and that $n\ge 3$).

\smallskip 
$\gamma(0)$:\quad We wish to prove that $u_0u_n\in E(G)$. 
We first observe that $N(u_0)\cap A\subseteq \{c_0,c_2\}$. Indeed, if $u_0$ would be adjacent to another vertex $c_{2i}\in A$, then there is a $c_{2i}$-claw with leaves $u_0,c_{2i-1},c_{2i+1}$. 
The vertex $u_0$ is adjacent to at least one of $c_0$ and $c_2$; otherwise there would be a $c_1$-claw with leaves $u_0,c_0,c_2$. If $u_0$ is adjacent to exactly one of them, then $u_0$ is already free in $I$, so $C$ can trivially be (internally) resolved in one move. 
So now we may assume that $N(u_0)\cap A=\{c_0,c_2\}$. 

If $m\le n$ then $u_n\in A$ so $u_n=c_0$, which shows that $u_0u_n\in E(G)$. 

Now suppose $m=n+1$, so $u_{n+1}=c_2$. Claim~$\alpha$ shows that $u_nc_2=u_nu_{n+1}\not\in E(G)$. Since $u_n\in L_0$, it holds that $u_nc_1\in E(G)$.
We conclude that $u_nc_0\in E(G)$, for otherwise there would be a $c_1$-claw with leaves $c_0,c_2,u_n$. 
Next, we note that $u_{n-1}c_0 \in E(G)$, because otherwise there would be a shorter path $u_0,\dots,u_{n-1},c_0$ in $D(G,C)$. Furthermore, $u_0 u_{n-1} \notin E(G)$ holds by $\delta(0)$, and $u_{n-1} u_n \notin E(G)$ by $\alpha$.
Combining these facts, we conclude that $u_0u_n\in E(G)$, because otherwise there would be a $c_0$-claw with leaves $u_0,u_n,u_{n-1}$ in $G$.

In the remaining case, $m\ge n+2$ holds. Then $u_nc_2\in E(G)$, for otherwise there would be a shorter path $u_0,\dots,u_n,c_2$ in $D(G,C)$.
In this case $u_0u_n\in E(G)$ follows since otherwise, there would be a $c_2$-claw with leaves $u_0,u_n,c_3$. This concludes the proof of Claim~$\gamma$ for the case $j=0$.

\smallskip 
$\delta(1)$:\quad By $\gamma(0)$, it holds that $u_0u_n\in E(G)$. 
Recall that $u_0u_1\not\in E(G)$, $c_{2n-1}u_1\not\in E(G)$ and $c_{2n-1}u_0\not\in E(G)$.
So $u_1u_n\not\in E(G)$, because otherwise there would be a $u_n$-claw with leaves $c_{-1},u_0,u_1$.

\smallskip 
$\delta(j)\implies \gamma(j)$ for $j=1,\dots,m-1$:\\
By $\delta(j)$, $u_j u_{j+n-1}\notin E(G)$ holds, and by $\alpha$, $u_{j+n}u_{j+n-1}\notin E(G)$ holds.
So $u_ju_{j+n}\in E(G)$, for otherwise there would be a $c_{2j-1}$-claw with leaves $u_j,u_{j+n},u_{j+n-1}$. 

\smallskip 
$\gamma(j)\implies \delta(j+1)$ for $j=1,\dots,m-1$:\\
We observe that $u_{j+n}u_{j-1}\in E(G)$, 
for otherwise there would be a shorter path $u_0,\dots,u_{j-1},u_{j+n},\dots,u_m$ in $D(G,C)$. 
Next, $u_{j+n}u_{j} \in E(G)$ by $\gamma(j)$, 
$u_{j-1}u_j\not\in E(G)$ and 
$u_j u_{j+1}\notin E(G)$ by $\alpha$, and 
$u_{j-1}u_{j+1} \notin E(G)$ 
by $\beta$, using that $n\ge 4$.
We conclude that $u_{j+1}u_{j+n}\not\in E(G)$, for otherwise there would be a $u_{j+n}$-claw with leaves $u_{j-1},u_j,u_{j+1}$.

\medskip
This concludes the induction proof of Claims~$\gamma$ and~$\delta$, and therefore the proof of the lemma.
\QED

\section{An Example of a Nontrivial Internal Resolving Sequence}

\begin{figure}[h!]
\caption{An example of a graph $G$, with an $I$-bad cycle of length 14, which is internally resolvable in $m=18$ steps. The vertices of $I$ are drawn as circles, and the other vertices of the $I$-bad cycle as squares. 
Half edges at the boundary of the figure continue on the other side. Vertices in each column $L_1,\dots,L_7$ form a clique. The directed path from the vertex $b$ to $I$ in $D(G,C)$ is also shown as a red dotted line to clarify the structure of $G$.}
\label{fig:internal_example_AGAIN}
\begin{tikzpicture}[xscale=1.6,yscale=0.2]
	\tikzstyle{ghost}=[inner sep=0pt,minimum size=0pt]
	\tikzstyle{v}=[circle,fill=black,draw=black!75,inner sep=0pt,minimum size=0.3em]
	\tikzstyle{I}=[circle,draw=black!75,inner sep=0pt,minimum size=0.8em]
	\tikzstyle{J}=[rectangle,draw=black!75,inner sep=0pt,minimum size=0.7em]
	\tikzstyle{splitl}=[path fading=west]
	\tikzstyle{splitr}=[path fading=east]
\pgfmathtruncatemacro\n{7}
\pgfmathtruncatemacro\m{18}
\pgfmathtruncatemacro\bi{1}
\pgfmathtruncatemacro\len{\m + \n - 1}
\foreach \j in {1,...,\len}
{
	\pgfmathtruncatemacro\x{mod(\j+\bi,\n)}
	\pgfmathtruncatemacro\npp{\n+2}
	\ifnum \j < \npp
		\node[v,label=-30:$u_{\j}$] (u\j) at (\x,\j) {};
	\else \ifnum \j < \m			
		\node[v] (u\j) at (\x,\j) {};
	\else
		\pgfmathtruncatemacro\iid{2*(mod(\j+\bi-2,\n)+1)}
		\node[I,label=above:$c_{\iid}$] (u\j) at (\x,\j) {};
	\fi \fi
}
\pgfmathtruncatemacro\x{mod(1+\bi,\n)}	
\node[v,label=below:$b$] (u0) at (\x-0.5,-2.5) {};
	
\foreach \j in {1,...,\len}
{
	\pgfmathtruncatemacro\x{mod(\j+\bi,\n)}
	\ifnum \x = 0
		\node[ghost] (ug\j) at (\n,\j) {};
	\fi
	\pgfmathtruncatemacro\x{mod(\j+\bi+1,\n)}
	\ifnum \x = 0
		\node[ghost] (ug\j) at (-1,\j) {};
	\fi
}

\foreach \j in {0,...,\len}
{
	
	\pgfmathtruncatemacro\jn{\j + \n}
	\pgfmathtruncatemacro\jnn{\j + 2*\n}
	\ifnum \jn<\m
		\foreach \k in {\jn,\jnn,...,\len}
		{
			\ifnum \j>0
				\draw[very thick] (u\j)--(u\k);
			\else
				\draw (u\j)--(u\k);
			\fi
		}
	\fi
	
	\pgfmathtruncatemacro\jnp{\j + \n + 1}
	\pgfmathtruncatemacro\jnnp{\j + 2*\n + 1}
	\pgfmathtruncatemacro\x{mod(\j+\bi+1,\n)}
	\ifnum \j<\len
		\foreach \k in {\jnp,\jnnp,...,\len}
		{
			\ifnum \k > \len
			\else
				\ifnum \x > 0 
					\draw (u\j)--(u\k);
				\else
					\draw[splitr] (u\j)--(ug\k);
					\draw[splitl] (ug\j)--(u\k);
				\fi
			\fi
		}		
	\fi
}

\foreach \i in {1,...,\n}
{
	\pgfmathtruncatemacro\iid{2*\i-1}
	\pgfmathtruncatemacro\x{mod(\i+\bi,\n)}
	\pgfmathtruncatemacro\y{\m+\n*1.5}
	\node[J,label=above:$c_{\iid}$] (c\iid) at (\x-0.5,\y) {};
	\ifnum \x=0
		\node[ghost] (cg\iid) at (\n-0.5,\y) {};
	\fi
}
\foreach \j in {0,...,\len}
{
	\pgfmathtruncatemacro\iid{2*mod(\j,\n)+1}
	\pgfmathtruncatemacro\x{mod(\j+\bi+1,\n)}
	\ifnum \x > 0
		\draw[gray] (u\j)--(c\iid);
	\else
		\draw[gray,splitl] (ug\j)--(c\iid);
		\draw[gray,splitr] (u\j)--(cg\iid);
	\fi

	\ifnum \j > 0
		\pgfmathtruncatemacro\iid{2*mod(\j-1,\n)+1}
		\pgfmathtruncatemacro\x{mod(\j+\bi,\n)}
			\draw[gray] (u\j)--(c\iid);
	\fi
}

\foreach \i in {1,...,\n}
{
	\pgfmathtruncatemacro\iid{2*\i-1}
	\pgfmathtruncatemacro\x{mod(\i+\bi,\n)}
	\node[ghost] (L\i) at (\x,-5) {$L_{\i}$};
	
}

\foreach \j in {1,...,\len}
{
	\pgfmathtruncatemacro\jm{\j-1}
	\pgfmathtruncatemacro\x{mod(\j+\bi,\n)}
	\ifnum \x > 0
		\draw[red,loosely dotted] (u\jm)--(u\j);
	\else
		\draw[splitr,red,loosely dotted] (u\jm)--(ug\j);
		\draw[splitl,red,loosely dotted] (ug\jm)--(u\j);
	\fi
}
\end{tikzpicture}

The graph $D(G,C)$ obtained from it:

\begin{tikzpicture}[xscale=1.6,yscale=0.2]
	\tikzstyle{ghost}=[inner sep=0pt,minimum size=0pt]
	\tikzstyle{v}=[circle,fill=black,draw=black!75,inner sep=0pt,minimum size=0.3em]
	\tikzstyle{I}=[circle,draw=black!75,inner sep=0pt,minimum size=0.8em]
	\tikzstyle{J}=[rectangle,draw=black!75,inner sep=0pt,minimum size=0.7em]
	\tikzstyle{splitl}=[path fading=west]
	\tikzstyle{splitr}=[path fading=east]

\pgfmathtruncatemacro\n{7} 
\pgfmathtruncatemacro\m{18} 
\pgfmathtruncatemacro\bi{1} 

\pgfmathtruncatemacro\len{\m + \n - 1}
\foreach \j in {1,...,\len}
{
	\pgfmathtruncatemacro\x{mod(\j+\bi,\n)}
	\pgfmathtruncatemacro\npp{\n+2}
	\ifnum \j < \npp
		\node[v,label=60:$u_{\j}$] (u\j) at (\x,\j) {};
	\else \ifnum \j < \m			
		\node[v,label=above:$u_{\j}$] (u\j) at (\x,\j) {};
	\else
		\pgfmathtruncatemacro\iid{2*(mod(\j+\bi-2,\n)+1)}
		\node[I,label=above:$u_{\j}$] (u\j) at (\x,\j) {};
	\fi \fi
}
\pgfmathtruncatemacro\x{mod(1+\bi,\n)}	
\node[v,label=below:$b$] (u0) at (\x-0.5,-2.5) {};
	
\foreach \j in {1,...,\len}
{
	\pgfmathtruncatemacro\x{mod(\j+\bi,\n)}
	\ifnum \x = 0
		\node[ghost] (ug\j) at (\n,\j) {};
	\fi
	\pgfmathtruncatemacro\x{mod(\j+\bi+1,\n)}
	\ifnum \x = 0
		\node[ghost] (ug\j) at (-1,\j) {};
	\fi
}

\foreach \j in {0,...,\len}
{
	
	\pgfmathtruncatemacro\jp{\j + 1}
	\pgfmathtruncatemacro\jnp{\j - \n + 1}
	\pgfmathtruncatemacro\x{mod(\j+\bi+1,\n)}
	\foreach \k in {\jp,\jnp,...,0}
	{
		\ifnum \k > 0
			\ifnum \k > \len
			\else
				\ifnum \x > 0 
					\draw[-latex'] (u\j)--(u\k);
				\else
					\draw[-latex',splitr] (u\j)--(ug\k);
					\draw[-latex',splitl] (ug\j)--(u\k);
				\fi
			\fi
		\fi
	}		
}

\foreach \i in {1,...,\n}
{
	\pgfmathtruncatemacro\iid{2*\i-1}
	\pgfmathtruncatemacro\x{mod(\i+\bi,\n)}
	\pgfmathtruncatemacro\y{\m+\n*1.5}
	\node[J,label=above:$c_{\iid}$] (c\iid) at (\x-0.5,\y) {};
	\ifnum \x=0
		\node[ghost] (cg\iid) at (\n-0.5,\y) {};
	\fi
}

\foreach \i in {1,...,\n}
{
	\pgfmathtruncatemacro\iid{2*\i-1}
	\pgfmathtruncatemacro\x{mod(\i+\bi,\n)}
	\node[ghost] (L\i) at (\x,-3) {$L_{\i}$};
}
\end{tikzpicture}
\end{figure}
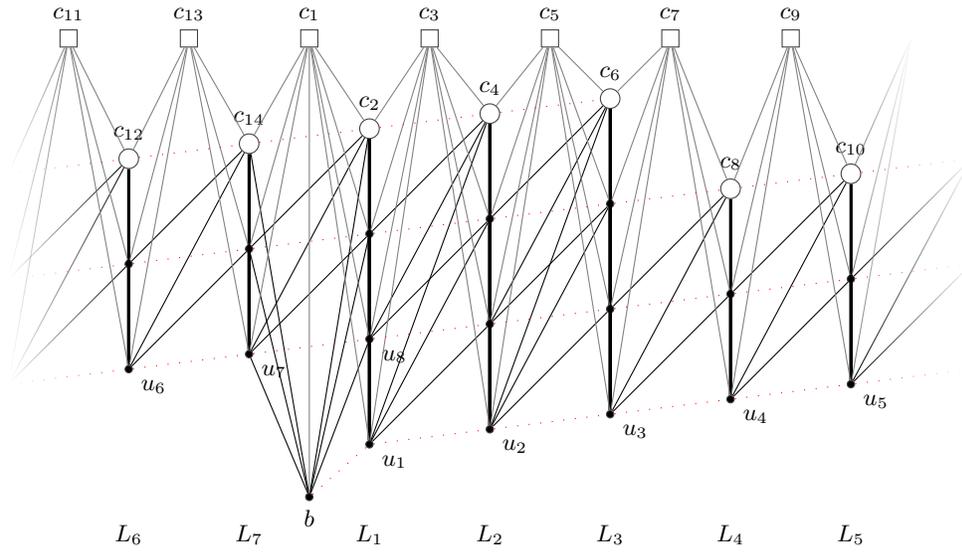

In Figure~\ref{fig:internal_example_AGAIN} on the next page, the construction of the graph $D(G,C)$ is illustrated. This figure shows an example where an elaborate \tsseq\ is required to (internally) resolve the given cycle.

\end{document}